\documentclass[1p]{elsarticle}

\usepackage{amsmath}
\usepackage{amssymb}
\usepackage{amsthm}

\usepackage{units}
\usepackage{hyperref}

\usepackage{tabularx}
\usepackage{comment}
\usepackage{float} 
\usepackage{lineno} 
\modulolinenumbers[5]

\journal{Safety Science}

\biboptions{authoryear}

\begin{document}

\begin{frontmatter}

\title{Modelling aerosol transport and virus exposure with numerical simulations in relation to SARS-CoV-2 transmission by inhalation indoors} 

\author[aaltoAddress]{V.~Vuorinen\corref{mycorrespondingauthor}}
\cortext[mycorrespondingauthor]{Corresponding author}
\ead{ville.vuorinen@aalto.fi}

\author[fmiAddress]{M.A.~Aarnio \fnref{myfootnote}}
\author[aaltoAddress6]{M.~Alava}
\author[aaltoAddress1]{V.~Alopaeus}
\author[fmiAddress,uhAddress1]{N.~Atanasova}
\author[fmiAddress]{M.~Auvinen}
\author[aaltoAddress3]{N.~Balasubramanian}

\author[aaltoAddress5]{H.~Bordbar}
\author[aaltoAddress4]{P.~Er\"ast\"o}

\author[aaltoAddress2]{R.~Grande}
\author[aaltoAddress3]{N.~Hayward}
\author[fmiAddress]{A.~Hellsten}
\author[aaltoAddress5]{S.~Hostikka}
\author[cscAddress]{J.~Hokkanen}
\author[aaltoAddress]{O.~Kaario}
\author[vttAddress]{A.~Karvinen}
\author[uhAddress]{I.~Kivist\"o}
\author[aaltoAddress6]{M.~Korhonen}
\author[aaltoAddress]{R.~Kosonen}
\author[essoteAddress]{J.~Kuusela}
\author[aaltoAddress]{S.~Lestinen}
\author[aaltoAddress]{E.~Laurila}
\author[aaltoAddress3]{H.J.~Nieminen}
\author[aaltoAddress]{P.~Peltonen}
\author[aaltoAddress1]{J.~Pokki}
\author[aaltoAddress6]{A.~Puisto}
\author[cscAddress]{P.~R\aa back}
\author[aaltoAddress6]{H.~Salmenjoki}
\author[uhAddress2,uhAddress3]{T.~Sironen}
\author[aaltoAddress2]{M.~\"Osterberg}

\fntext[myfootnote]{Co-authors are listed in alphabetical order, not based on the extent of their contributions.}

\address[aaltoAddress]{Department of Mechanical Engineering, Aalto University, FI-00076 AALTO, Finland}
\address[fmiAddress]{Atmospheric Dispersion Modelling, Atmospheric Composition Research, Finnish Meteorological Institute, FI-00101, Helsinki, Finland}
\address[aaltoAddress1]{Department of Chemical and Metallurgical Engineering, Aalto University, FI-00076 AALTO, Finland}
\address[aaltoAddress2]{Department of Bioproducts and Biosystems, Aalto University, FI-00076 AALTO, Finland}
\address[aaltoAddress3]{Department of Neuroscience and Biomedical Engineering, Aalto University, FI-00076 AALTO, Finland}
\address[aaltoAddress4]{Department of Information and Service Management, Aalto University, FI-00076 AALTO, Finland}
\address[aaltoAddress5]{Department of Civil Engineering, Aalto University, FI-00076 AALTO, Finland}
\address[aaltoAddress6]{Department of Applied Physics, Aalto University, FI-00076 AALTO, Finland}
\address[uhAddress1]{Molecular and Integrative Biosciences Research Programme, Faculty of Biological and Environmental Sciences, University of Helsinki, Finland}
\address[uhAddress2]{Department of Virology, Faculty of Medicine, University of Helsinki, Helsinki, Finland}
\address[uhAddress3]{Department of Veterinary Biosciences, Faculty of Veterinary Medicine, University of Helsinki, Helsinki, Finland}
\address[vttAddress]{VTT Technical Research Centre of Finland Ltd, Finland}
\address[cscAddress]{CSC-IT Center for Science Ltd, FI-02101, Finland}
\address[essoteAddress]{Emergency Department, 
Mikkeli Central Hospital, The South Savo Social and Health Care Authority, FI-50100, Finland}
\begin{abstract}
We provide research findings on the physics of aerosol and droplet dispersion relevant to the hypothesized aerosol transmission of SARS-CoV-2 during the current pandemic. We utilize physics-based modeling at different levels of complexity, along with previous literature on coronaviruses, to investigate the possibility of airborne transmission. 
The previous literature, our 0D-3D simulations by various physics-based models, and theoretical calculations, indicate that the typical size range of speech and cough originated droplets ($d\leq 20~\unit{\mu m}$) allows lingering in the air for $O(1~\unit{h}$) so that they could be inhaled. Consistent with the previous literature, numerical evidence on the rapid drying process of even large droplets, up to sizes $O(100~\unit{\mu m})$, into  droplet nuclei/aerosols is provided. Based on the literature and the public media sources, we provide evidence that the individuals, who have been tested positive on COVID-19, could have been exposed to aerosols/droplet nuclei by inhaling them in significant numbers e.g. $O(100)$. By 3D scale-resolving computational fluid dynamics (CFD) simulations, we give various examples on the transport and dilution of aerosols ($d\leq 20~\unit{\mu m}$) over distances $O(10~\unit{m})$ in generic environments. We study susceptible and infected individuals in generic public places by Monte-Carlo modelling. The developed model takes into account the locally varying aerosol concentration levels which the susceptible accumulate via inhalation. The introduced concept, 'exposure time' to virus containing aerosols is proposed to complement the traditional 'safety distance' thinking. We show that the exposure time to inhale $O(100)$ aerosols could range from $O(1~\unit{s})$ to $O(1~\unit{min})$ or even to $O(1~\unit{h})$ depending on the situation. The Monte-Carlo simulations, along with the theory, provide clear quantitative insight to the exposure time in different public indoor environments. 
\end{abstract}


\end{frontmatter}


\section{Introduction}
\label{sec:intro}

\subsection*{Background}
COVID-19 disease is caused by the novel SARS-CoV-2 coronavirus, with over 5 million known infections by May 2020 worldwide.
The disease severity ranges from asymptomatic to life threatening illness, with over 300,000 deaths reported globally to date. For SARS-CoV-2, transmission by large droplets, good hand hygiene and social distancing have been mostly emphasized during the first quarter of 2020 \citep{CDC2020_socialDistancing,WHO2020_adviceForPublic}. During April 2020, the possibility for airborne transmission was discussed in various research communications providing insightful argumentation on the possibility of airborne transmission with common linkage to pre-symptomatic cases \citep{Asadi2020,Morawska2020,Anderson2020}. The relative importance of the transmission mechanisms for different viruses is an open research question \citep{Kutter2018}.

For SARS-CoV-2, a recent article quantifies the relative contributions to the basic reproductive number $R_0$ during the early phase of the epidemic in China \citep{Ferretti2020}. The article reports transmission from individuals who are pre-symptomatic (46\%), symptomatic (32\%) or asymptomatic (10\%), and from the environment (6\%) with the latter two lacking confirmation. Such a transmission refers to droplets that are initially small or larger droplets that dry almost immediately outside the mouth into light/small droplet nuclei. These droplet nuclei can linger in the air similar to aerosols or other sufficiently small particles \citep{Asadi2019,Asadi2020,Stadnytskyi2020}.

In general, it is well understood that transmission by close contact plays an important role in the spreading of infectious diseases \citep{Halloran2010}. In particular, infection risk is affected by the number of contacts along with other factors such as the human-human distance or the type and duration of contact itself \citep{DeCao2014}. This is well acknowledged for the pandemic influenza \citep{Haber2007} and specifically in the case of varicella and parvovirus \citep{Melegaro2011}, as well as for SARS-CoV-1 \citep{Riley2003}. Coronaviruses are known as important veterinary pathogens and are also a cause of the common cold in people. In the last 20 years, the picture has changed as three major outbreaks of severe respiratory disease have been caused by coronaviruses. In 2002 and 2003, a global outbreak of severe acute respiratory syndrome (SARS) drew attention to the pandemic potential of novel coronaviruses. SARS caused 8096 known cases and 774 deaths before the outbreak was controlled in less than a year through vigorous quarantine measures. Ten years later, Middle East respiratory syndrome (MERS) emerged, causing even more concern due to its high case fatality rate. MERS is not efficiently transmitted from person to person, and the case count has remained low \citep{daCosta2020}. After these two rather limited outbreaks, SARS-CoV-2 has surprised the world again with the high transmissibility of the pathogen. 

International guidelines have been established for promoting the safety of public indoor spaces during the present pandemic. Previous evidence for SARS-CoV-1 indicates that virus-laden aerosols could be transmitted within ventilated spaces \citep{Li2007}. In fact, international organizations such as the Federation of European Heating, Ventilation and Air Conditioning Associations (REHVA) and The American Society of Heating, Refrigerating and Air-Conditioning Engineers (ASHRAE) have recently considered the possibility of the airborne spread of SARS-CoV-2 \citep{REHVA2020,ASHRAE2020}.

This study brings to focus one of the least understood transmission mechanisms of SARS-CoV-2: transmission by inhalation of virus-containing aerosols. While it may not be the principal transmission mode, it constitutes an essential entity which is needed to formulate a complete picture on the epidemiology of the COVID-19. The topic emphasizes the importance of fluid flow physics, as highlighted by a recent review article by \citet{mittal_ni_seo_2020} summarizing, for instance, the fundamentals on the formation of exhaled droplets and their subsequent drying and evaporation processes. 

This work makes a multidisciplinary contribution to the subject matter. We discuss what is known on the SARS-CoV-2 virus, the physical processes related to the aerosolization of the exhaled droplets and, using high fidelity Computational Fluid Dynamics (CFD) approach, we examine in unprecedented detail a high-risk scenario where an infected individual coughs within a public indoor space. In addition, this work seeks to fill the gap between the local aerosol release due to speaking or coughing, and the exposure risk of other people sharing the space by incorporating the motion of the people. This is achieved by utilizing dispersion results from the CFD-modelling in Monte-Carlo simulations. To our knowledge, this kind of combined modelling study has never been published in the context of exposure to virus-laden aerosols.

In earlier studies, pathogen transmission has been investigated widely in various indoor environments \citep{Li2007,Morawska2017,Ai2018}. In many cases, CFD and experimental studies have been conducted to examine the airborne transmission in simplified environments \citep{Yu2004,Zhang2006,Liu2017} that may have an association with ventilation \citep{Li2005,Olmedo2012,Cao2015}, airflow interaction \citep{Bivolarova2017,Li2018a,Ai2019}, in hospitals \citep{Yu2005,King2015,Cho2019} and offices \citep{He2011}. Certain studies also consider transport induced by human motion \citep{Edge2005,Han2014} as well as breathing, talking and coughing \citep{Gupta2010,Licina2015,Li2018b}.
Coughing has been simulated previously using different turbulence modelling approaches with Lagrangian particles. For example, coughing in an aircraft cabin has been simulated by \citet{Wan2009}. The researchers validated their simulation results against the experimental results of \citet{Sze2009}. In their experiments and simulations, realistic thermal loads produced by passengers along with standard ventilation conditions found in passenger aircraft were used. 
One of the first papers where CFD simulation was used to investigate the trajectories of airborne particles in this context, was by \citet{Seymour2000} where hospital isolating room applications were investigated.

Most of the previous studies are based on the Reynolds averaged Navier-Stokes approach (RANS) in which turbulent motion is not explicitly resolved and its effects on the mean flow field and aerosol dispersion are modelled using some approximate turbulence closure model involving significant uncertainties. Indoor ventilation air flows are typically dominated by turbulent motion while the influence of the mean flow on the dispersion and dilution of an aerosol cloud may be weak compared with the influence of the turbulent motion. Therefore, we claim that it is important to explicitly resolve at least the most influential part of the turbulent motion. In this study we achieve this by employing high-resolution Large-Eddy Simulation (LES). In our view, LES is the most suitable approach to the present problem although it is computationally more expensive than RANS. 
Earlier, only few LES-studies with some relevance to the present study have been published \citep{Tian2007, Berrouk2010, Liu2012}. 

\subsection*{Particular research gaps}

Based on the literature, the following research gaps are identified. 1) Synthesizing the literature based information on the size ranges and the numbers of cough or speech generated droplets is necessary in order to form a complete picture on the relevance of airborne transmission. Such information is also necessary input for e.g. CFD  simulations and post-processing the results. 2) There is ambiguity about the definitions of aerosols/airborne droplets/droplet nuclei in the literature. The ambiguity must be removed in order to form a complete picture on airborne transmission and to understand the dependency between droplet size and the time those droplets could linger in the air. 3) The exposure to inhaled aerosol particle counts has not been estimated previously in the reported COVID-19 cases. 4) It is clear that aerosols transport over long distances with air flow but there is a lack of understanding on the connection between the physical distance and the exposure time i.e. the time during which one could accumulate a critical dose by inhalation. 5) There is a demand to better understand transmission of SARS-CoV-2 in public places under various circumstances. 6) Quantitative decision making metrics, to assess safety of public premises, are needed.     

\subsection*{Objectives}
The research hypothesis is the possibility of airborne transmission of SARS-CoV-2. The objectives are set to fill the research gaps (see above). 
The main objectives of the present study are formulated as follows.
\begin{enumerate}
    \item Synthesize the existing knowledge on the role of coughing, speaking and breathing in the formation of aerosols and small droplets. Specific attention is given to information concerning the number, size distribution, and rate of production of particles released while coughing or speaking. 
    \item Refine the concept of airborne droplet based on simulations and previously known information. Use 0D-3D numerical modeling to discuss the connection between droplet size, droplet sedimentation time and the time of evaporation. 
    \item Use literature data on speech generated aerosol production rates (see item~1) and propose estimates for the aerosol exposure during distinct gatherings where SARS-CoV-2 transmission has been reported to occur.
    \item Demonstrate aerosol transport over long distances using 3D CFD modelling and the LES approach. Determine the characteristic, generalized time scale for cough-released aerosol cloud dilution. 
    \item Define an analysis methodology for the CFD results to determine and quantify the spatial evolution and extent of a high risk zone in the vicinity of a cough plume.
    \item Characterize certain risk scenarios for transmission via inhalation in public premises as a function of different percentages of infectious individuals and different person-number densities.
\end{enumerate}

\subsection*{Outline}  

\begin{figure}[H]
	\centering
	\makebox[0pt]{
		\includegraphics[width=12cm]{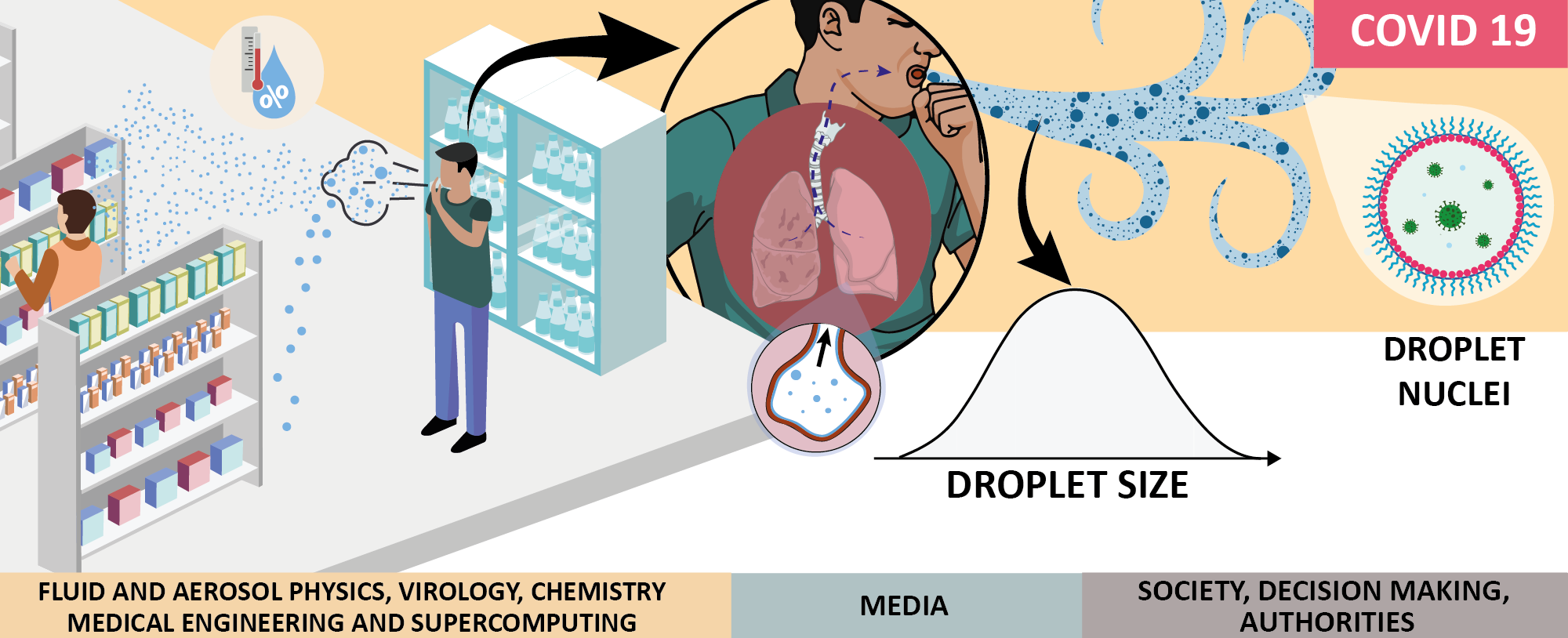}
	}
	\caption{Outline of the present work. The multidisciplinary consortium investigates the possibility of aerosol based transmission of SARS-CoV-2 virus in public places by quantifying the exposure to cough-released aerosols. The consortium approaches the topic from 0D-3D physics based simulation viewpoint, along with insights from virology and medical literature with relevance to coronaviruses and droplets of respiratory origin. }
	\label{fig:GraphicalAbstract2}
\end{figure}

This report documents the research project (see Fig.~\ref{fig:GraphicalAbstract2}) undertaken by a multidisciplinary research consortium which was established on March 21st 2020 between two Finnish universities, two state research organizations and a super-computing center. The consortium consists of numerical modelling experts in flow physics, spray and aerosol physics, medical physics, engineering, social network dynamics, virology, chemists, and medical doctors. We explore the emerging hypothesis on the possibility of airborne transmission of SARS-CoV-2.

%

This paper is organized as follows: General definitions are first declared in Section~\ref{sec:definitions}. In Section~\ref{sec:virology}, we review what is known on the SARS CoV-2 virus at present. Section~\ref{sec:particleDistribution} consists of a literature survey on droplet size distributions of respiratory origin. In Section~\ref{sec:droplets}, certain droplet size effects, including sedimentation, evaporation and the effect of surfactants, on droplet properties are recapped using 0D-1D numerical simulations. The applied 3D CFD and 2D Monte-Carlo methods are laid out in Section~\ref{sec:models} while the respective simulation results are analysed and discussed in Section~\ref{sec:results}. Conclusions, recommendations, and suggestions for future research topics are discussed in Section~\ref{sec:conclusions}.


\section{Definitions}
\label{sec:definitions}

There has been great ambiguity in the definitions of the most essential concepts related to potential disease transmission routes, starting from the concepts of "airborne transmission" and "droplet transmission", along with a dispute of what is considered as an "aerosol", "small droplet", or "large droplet". This ambiguity has resulted in severe misunderstanding between researchers, and in the worst case scenario, research findings and disciplines have been disconnected. Even in WHO's own publications, contradictory definitions exist, see e.g. \citet[Annex~C]{WHO2009_ventilation} and \citet{WHO2020_transmissionModes}. Throughout this paper, we use the following definitions:
\begin{itemize}
\item \textbf{Droplet size:} A highly dynamic quantity which typically reduces very rapidly when a liquid droplet is exhaled. E.g. water droplets with diameter $d\leq 50~\unit{\mu m}$, would evaporate at relative humidity RH = 50\% in less than 3 seconds into water vapor. Since the sedimentation time of the respective solid particles (assuming 1.6m initial height and still air) would be about $30~\unit{s}$, the respective water droplets would evaporate completely before reaching the ground. For mucus, such a drying process would yield left-over droplet nuclei which could potentially carry viruses. Hence, there is no fixed droplet size. Larger droplets than typically thought, e.g. $O(100~\unit{\mu m})$ in initial size, may become aerosol particles because of rapid drying. 

\item \textbf{Aerosol:} May refer to either 1) an aerosol particle or 2) a suspension of liquid droplets, droplet nuclei or particles in air. Aerosol particles in the present context may contain infectious pathogens, epithelial and other cells or remains of those, natural electrolytes and other substances from mucus and saliva, and water typically evaporating rapidly depending on the relative humidity of the surrounding air. Aerosol particles remain in the air for long enough for them to be inhaled and they can be transferred over long distances by indoor air flow. For drying liquid droplets (e.g. $d\leq 50~\unit{\mu m}$), a several minute time of remaining 'airborne' is numerically demonstrated in still ambient air while ambient flow can sustain those droplets for much longer times in the air. For example, in the CFD simulations herein, solid particles with $d = 20~\unit{\mu m}$ having density of water are demonstrated to behave as aerosols lingering at least several minutes in the turbulent air flow in conditions relevant to public places.

\item \textbf{Droplet nuclei:} Residuals of droplets after drying to moisture in equilibrium with ambient air. In addition to equilibrium moisture, droplet nuclei contain non-volatile substances from the original wet droplet. Dynamical transport of droplet nuclei is dictated mainly by external air flow unless the released wet droplets were large enough to sediment before drying, i.e. droplet nuclei are aerosols. If droplets sediment before drying, dried-out residuals on surfaces are not considered droplet nuclei.

\item \textbf{Small droplet:} Droplets which stay airborne for significant enough times (e.g. tens of seconds, minutes) to be inhaled. These droplets dry into droplet nuclei very rapidly. Precise size cutoff is affected by the ambient conditions, in particular ambient ventilation and air flow patterns. For example, under still ambient conditions $d\leq 100~\unit{\mu m}$ droplets could be considered to be small. For more discussion about the size limits for a small droplet, see \citet{Xie2007}.

\item \textbf{Large droplet:} Droplets which stay airborne only for short times (e.g. seconds). These droplets could be e.g. larger than $200~\unit{\mu m}$. Movement of large droplets is mainly determined by gravitational settling and to less extent by the ambient air flow. Drying rate is typically slow enough for the large droplets to settle to the ground or other surrounding surfaces before becoming an aerosol.

\item \textbf{Droplet transmission:} 
Literature definitions are contradictory and, in practice, terminology varies from person to person. The term refers to virus transmitted via droplets produced by an infected human, via sneezing, coughing, speaking, or other means either by respiratory droplets of initial size $d>5~\unit{\mu m}$ or fine aerosols $<5~\unit{\mu m}$ \citep{Tellier2019,Asadi2020}. Airborne transmission, by inhalation of airborne droplets, can be considered to be a droplet transmission. 

\item \textbf{Airborne transmission:} Any pathogen that can be transmitted via air as e.g. aerosols, liquid droplets, or dust. Any droplet size that can potentially be carried by air, from the infection source to the susceptible person, may cause a potential vector for the airborne route. Transmission by inhalation of droplets or aerosols belongs to this category as well. 

\item \textbf{Transmission by inhalation:} Transmission by inhaling aerosols, whether they may be droplets, droplet nuclei or other viral particles. 
\end{itemize}

\section{Virology of SARS-CoV-2}
\label{sec:virology}

\subsection{Background context}

SARS-CoV-2 transmission is thought to be mostly by inhalation of droplets and contact contamination \citep{Jin2000}. The pathogenesis has been proposed to start with cell entry at the respiratory tract via angiotensin-converting enzyme 2 (ACE2) receptor proteins \citep{Li2003}. Viral replication leads to local and systemic inflammatory responses, resulting in a cytokine storm that promotes variable and severe pathological responses \citep{Chung2020} and multi-organ dysfunction \citep{Jin2000}. There is currently no proven pharmacological intervention for COVID-19, thus treatment is largely supportive and often requires intensive care.

Coronaviruses are a group of viruses that infect diverse vertebrates including humans. Such viruses have non-segmented, positive sense single-stranded RNA genomes of approximately 27 to 30~kb, which are the largest known RNA virus genomes \citep{Woo2009}. The SARS-CoV-2 virus has been reported to be of size 80-100~nm being approximately of size by factor 10-20 smaller in diameter than commonly observed droplets from speaking \citep{Park2019}. Coronaviruses are divided into four genera with different and diverse host animals: viruses of Alphacoronavirus and Betacoronavirus genera infect mammalian hosts, while viruses of Gammacoronavirus and Deltacoronavirus genera mainly infect avian hosts \citep{ICTV2020}. To date, seven coronaviruses are known to infect humans \citep{Chen2020}. These are coronaviruses 229E (subgenus Duvinacovirus of genus Alphacoronavirus), NL63 (subgenus Setracovirus of genus Alphacoronavirus), OC43 and HKU1 (subgenus Embecovirus of genus Betacoronavirus), all that cause mild seasonal upper respiratory infections, and MERS-CoV (subgenus Merbecovirus of genus Betacoronavirus), SARS-CoV (subgenus Sarbecovirus of genus Betacoronavirus) and SARS-CoV-2 with more serious disease outcomes. Compared to the other human coronaviruses, SARS-CoV-2 is genetically most similar to SARS-CoV with approximately 80\% RNA genome sequence identity, but it is even more closely related ($\sim$85\%) to bat coronavirus strains ZC45 and ZXC21 \citep{Zhu2020}, indicating a potential origin from bats.

COVID-19 patients can spread the disease before the onset of clinical symptoms and asymptomatic persons infected with the SARS-CoV-2 virus are able to transmit the disease, which makes it challenging to control the progression of the pandemic \citep{Wang2020,Rothe2020}. Asymptomatic cases have also been reported for SARS and MERS coronaviruses \citep{Oboho2015,Che2006}.

\subsection{Symptoms and detected viral loads in COVID-19 patients}

Symptoms of COVID-19 include fever, cough, myalgia or fatigue, sputum production, headache, coughing up blood, diarrhoea and shortness of breath \citep{Huang2020}. The severity of the disease ranges from asymptomatic to fatal, and the mortality rate is high for the elderly and for those with pre-existing comorbidities. The viral loads, as identified by cycle threshold (Ct) values of RT-PCR assay, have been reported to be highest soon after the onset of symptoms reaching up to 10$^8$ copies per ml, and typically be below detection threshold by 5 to 18 days since onset of symptoms, with asymptomatic patients having a similar pattern \citep{Zou2020}. Infectious virus can be isolated from samples derived from either the throat or lungs, and shedding of viral RNA can also outlast the symptoms \citep{Wolfel2020}. Viral load and persistence are both associated with the severity of the disease \citep{Zheng2020}. Significant individual variation in viral shedding has been noted, and the fraction of high emitting individuals may be as high as 10\%. 

\subsection{Transmission modes}

Viruses that cause respiratory illnesses in humans often have multiple modes of transmission. Such viruses may spread by droplet transmission via either $>5~\unit{\mu m}$ respiratory droplets or fine, $< 5~\unit{\mu m}$ aerosol droplets, by direct contact such as hand shake or indirect e.g. via contaminated surfaces \citep{Tellier2019,Asadi2020}. Seasonal coronaviruses (e.g. strains NL63, OC43, 229E and HKU1), influenza viruses (A and B), and rhinoviruses have all been shown to be transmitted by respiratory droplets, both by close contact or via aerosols \citep{Leung2020}. Recent study demonstrated that the virus can remain infective in aerosols for 3~h and on surfaces for 72~h in laboratory conditions \citep{Doremalen2020}. However, evidence about acquiring the infection via contaminated surfaces is scarce. Infectious human seasonal coronavirus 229E was detected after 6~d from aerosols at 20$^\circ$C and 50\% RH, having a half-life of 70~h in the laboratory \citep{Ijaz1985}. The recommended safety precautions, such as washing hands and keeping a safety distance can give reasonable protection against surface contamination and close-contact droplet transmission. The infectivity of SARS-CoV-2 in micron-sized aerosol particles has not yet been demonstrated. 


As a matter of fact, aerosol transmission is not well understood for any virus. Aerosol particle size distribution versus the number of infectious viruses within one particle, is particularly weakly understood. One study detected on average 142 plaque forming units (PFU) of infectious influenza A viruses in $0.3-8~\unit{\mu m}$ aerosol particles produced during six coughs \citep{Lindsley2015}. Another study reported influenza A transmission between ferrets to be mediated by aerosol particles of $>1.5~\unit{\mu m}$ in size \citep{Zhou2018}. Interestingly, certain genetic determinants and the hemagglutinin–neuraminidase balance of influenza viruses have been shown to increase the ability of the virus to spread in aerosol droplets in ferrets \citep{Yen2011}. In one report, SARS-CoV-2 RNA was detected from aerosol particles $>4~\unit{\mu m}$ and $1-4~\unit{\mu m}$ from air samples collected from two patient hospital rooms in Singapore \citep{Chia2020}. SARS-CoV-2 is genetically closely related to SARS-CoV-1 for which reports about potential aerosol transmission exist \citep{Yu2004}.

\section{Droplet size distribution}
\label{sec:particleDistribution}

The aerosol particles in the exhalation air by an infected person during breathing, speaking, sneezing, and coughing may carry airborne pathogens which may cause infection diseases if inhaled by others \citep{Edwards2004}. Various factors and physical mechanisms, including the site of origin, and reopening of small airways in the respiratory system, will affect the particle properties in the exhaled breath \citep{Bake2019}. The size of these particles may vary from $10~\unit{nm}$ to $100~\unit{\mu m}$ and above. The particles are droplet-like i.e. wet when entering the mouth, but will dry, forming dry particles when exhaled and mixed with the ambient air. The drying rate depends on the ambient relative humidity as well as on the particle diameter, surfactants along with particle properties (e.g. proteins in the mucus). 

While very large droplets follow a ballistic trajectory and impact on or fall onto surfaces within a limited radius of the source, the particles with intermediate size, e.g. $10-100~\unit{\mu m}$, can travel longer distances \citep{Lindsley2013}. However, the small particles ($<10~\unit{\mu m}$) least likely impact and settle on to surfaces and can float on the air and spread much further following the air flow stream especially after being dried \citep{Lindsley2013}. Moreover, the wet particles can dry very rapidly and transform to dry aerosol. For instance, the drying times for $100~\unit{\mu m}$ and $50~\unit{\mu m}$ droplets in air at 50\% relative humidity are reported to be 1.3 and 0.3~s, respectively \citep{Lenhart2004}. On complete evaporation, the particles may be small enough to remain airborne in the indoor air flow. These small aerosol particles could potentially carry viruses and also contribute to the spreading of the epidemic as many of them are still large enough to contain thousands of viral and bacterial pathogens \citep{ACGIH1999} with usual size of $\sim 25~\unit{nm}$ to $5~\unit{\mu m}$ \citep{Edwards2004}.

In a recent study \citep{Leung2020} on the efficacy of facial mask in reducing the risk of coronavirus transmission via droplets and aerosols, the viral RNA was detected in 30\% of the larger droplets ($>5~\unit{\mu m}$) while it was detected in 40\% of the smaller aerosols ($<5~\unit{\mu m}$). It is also recently argued that the small aerosols ($<5~\unit{\mu m}$) exhaled in normal speech plausibly serve as an important and under-recognized transmission agent for SARS-CoV-2 \citep{Asadi2020}.

\begin{table}[!t]
	\centering
	\caption{Droplet size distributions of respiratory origin (partly reproduced from \citep{Han2013} }
	\label{tab:tab1}
	\makebox[0pt]{
    	\begin{tabular}{p{4cm}p{3cm}p{3cm}p{5cm}}
    		\hline
    		Reference & Topic & Respiratory origin & Main result \\ \hline
    		
    		\citet{Yang2007} & 54 healthy subjects & cough & size range from 0.62 to 15.9 $\mu$m with the average mode of 8.35 $\mu$m\\
    		
    		\citet{Fennelly2004} & 16 patients infected with tuberculosis & cough & most particles were in the respirable size range\\
    		
    		\citet{PAPINENI1997} & five healthy subjects & cough, mouth breathing, nose breathing, talking & 87 \%, 86 \%, 88 \%, and 84 \% of the particles had diameters of less than 1 $\mu$m for cough, mouth breathing, nose breathing and talking, respectively.\\

    		\citet{Asadi2019} & 10-30 healthy subjects & speaking, vocalization, breathing & Positive correlation between the rate of particle emission and the loudness. 1 to 50 particles per second (0.06 to 3 particles per cm3). \\

    		\citet{Edwards2004} & 12 patients infected with influenza & breath & size range from 0.15 to 0.19 $\mu$m \\
    		
    		\citet{Fabian2008} & 12 patients infected with influenza & breath & over 87\% of particles exhaled were under 1 $\mu$m in diameter \\
    		
    		\citet{Fabian2011} & three healthy subjects and 16 patients infected with human rhinovirus & breath & 82\% of particles detected were 0.300–0.499 $\mu$m. \\ \hline
    	\end{tabular}
	}
\end{table}

A vast number of studies have focused on the characterization of particle size distributions exhaled by healthy and infected subjects in various modes including talk, cough and sneeze of which some are summarized in Table~\ref{tab:tab1}. Unfortunately, there is remarkable inconsistency in the reported data which can be traced back to the used particle detection methods \citep{Han2013}. Moreover, the size distribution of sneeze originated droplets is also largely affected by individual response to sneezing.

Using a laser particle size analyzer, \citet{Han2013} measured the volume-based size distributions of sneeze droplets at the mouth recognizing uni-modal and bi-modal patterns. Studying the coughing of influenza patients by using a laser aerosol particle spectrometer, it is found that the individuals with influenza cough out a greater volume of aerosol particles than they do when they are healthy \citep{Lindsley2012}.

Experimental measurements of particle size distribution of respiratory particles showed that the log-normal distribution is a good approximation \citep{Bake2019}. Hence, the particle size distribution is usually described with three parameters of geometric mean diameter, geometric standard deviation and total number or mass concentration \citep{Bake2019}. For instance, the measurements of \citet{Lindsley2012} showed that the average cough aerosol volume of patients with influenza is 38.3 picoliters~(pL) of particles per cough with standard deviation (SD) of 43.7; after patients recovered, the average volume was $26.4~\unit{pL}$ per cough (SD 45.6). The number of particles produced per cough was also higher when subjects had influenza (average 75,400 particles/cough, SD 97,300) compared with afterward (average 52,200, SD 98,600) \citep{Lindsley2012}. 

In another experimental study, the interferometric Mie imaging (IMI) technique together with the particle image velocimetry (PIV) technique were used to characterize the droplet size distributions during coughing and speaking immediately at the mouth opening which made the effect of evaporation and condensation negligible \citep{Chao2009}. For a healthy person, the geometric mean diameter of droplets from coughing and speaking was measured as $13.5~\unit{\mu m}$ and $16.0~\unit{\mu m}$, respectively while the total number of droplets expelled ranged from 947 to 2085 per cough and 112–6720 for speaking \citep{Chao2009}.In the latter case, the number of particles are counted while the subjects are loudly counting from 1 to 100 \citep{Loudon}. The estimated droplet concentrations for coughing ranged from 2.4 to $5.2~\unit{cm^{-3}}$ per cough and from 0.004 to $0.223~\unit{cm^{-3}}$ for speaking \citep{Chao2009}. 

While the role of coughing and sneezing in spreading the virus contained aerosols and droplets is often emphasized, the number of particles produced by speaking is also significant especially as it is normally done continuously over a longer period of time. The effect of loudness on the number of emitted particles is also substantial. \citet{Asadi2019} reported that the rate of particle emission during normal human speech exhibits positive correlation with the loudness (amplitude) of vocalization, ranging from approximately 1 to 50 particles per second (0.06 to 3 particles per $\unit{cm^3}$) for low to high amplitudes. It is inline with another recent experimental observation done by laser light scattering showing that the number of droplet flashes increases with the loudness of speech \citep{Anfinrud}. It is also observed that some individuals (aka speech super-emitters) emit particles at a rate more than an order of magnitude larger than their peers \citep{Asadi2019}. 

Moreover, the viscoelastic properties of mucus have been found to substantially affect the size distribution and number of droplets generated during coughing \citep{AnwarulHasan2010}. It is also noticeable that normal mouth breathing has been reported to produce airborne droplets in larger numbers than in coughing, nose breathing, or talking \citep{PAPINENI1997}. 

As briefly reviewed, there is quite a lot of inconsistency in the experimental data reported for particle number and size distribution of various respiratory behaviors including speaking, singing, coughing and sneezing. Many of the previous results may be biased to the measurement methods and dynamical effects due to e.g. rapid drying of droplets. For example, a recent study reports very high intensity of particle number production during speech \citep{Stadnytskyi2020}. 
After all, in the CFD part of the present study, we investigate a scenario where a single cough releases 40,000 particles of 10 and $20~\unit{\mu m}$ size being consistent with the size range and number of particles reported before, see for instance \citet{Lindsley2012}. For the Monte-Carlo modeling, we investigate a speaking person emitting at the rate of 5 particles/second \citep{Asadi2019} while in a coughing scenario the same 40,000 particles per cough is assumed. 

\section{Droplet size effects}
\label{sec:droplets}

\subsection{Sedimentation and evaporation - stagnant air flow}
The aim of this section is to characterize the difference between small and large droplets. In particular, to understand which droplets can stay airborne for long enough to be inhaled. We note that there is still a rather great confusion regarding the droplet sizes and difference between 'aerosol' and 'droplet' during the COVID-19 pandemic. As mentioned, droplets are never of constant size because they evaporate rapidly. Next, we briefly recap and characterize the concepts of droplet sedimentation time and evaporation time.

We use standard numerical methods to solve the droplet equation of motion, time-varying temperature, and evaporation for water droplets of initial diameter $1\leq d \leq 200~\unit{\mu m}$ \citep{Xie2007}. We investigate droplets falling freely in still air from the height $h=1.625~\unit{m}$ following their equation of motion under gravity $g=9.81~\unit{m\,s^{-2}}$, and droplet kinematic timescale $\tau_p = \frac{\rho_p d^2}{18\nu_g \rho_g}$. Under steady state, assuming stagnant ambient air, the droplet terminal velocity $v_p = \tau_p g$ yielding the sedimentation time $t_s = h / v_p$. This indicates that e.g.~solid particles ($\rho_p = 1000\unit{kg/m^3}$) of size $10~\unit{\mu m}$ would settle to the ground in about $8$ minutes while the respective time would be approximately $2$ minutes for $20~\unit{\mu m}$ particles. Hence, droplet size is essential in distinguishing between an airborne droplet, which can be suspended in air and inhaled, and a falling, ballistic droplet which meets the ground or surfaces within seconds. We note that the given definition of $t_s$ holds in particular for stationary air and non-evaporating droplets. In practice, turbulent flow transports droplets upwards and downwards keeping them suspended in air for sometimes longer, sometimes shorter period of time than predicted by $t_s$. 

\begin{figure}[H]
	\centering
	\makebox[0pt]{
		\includegraphics[width=1.0\linewidth]{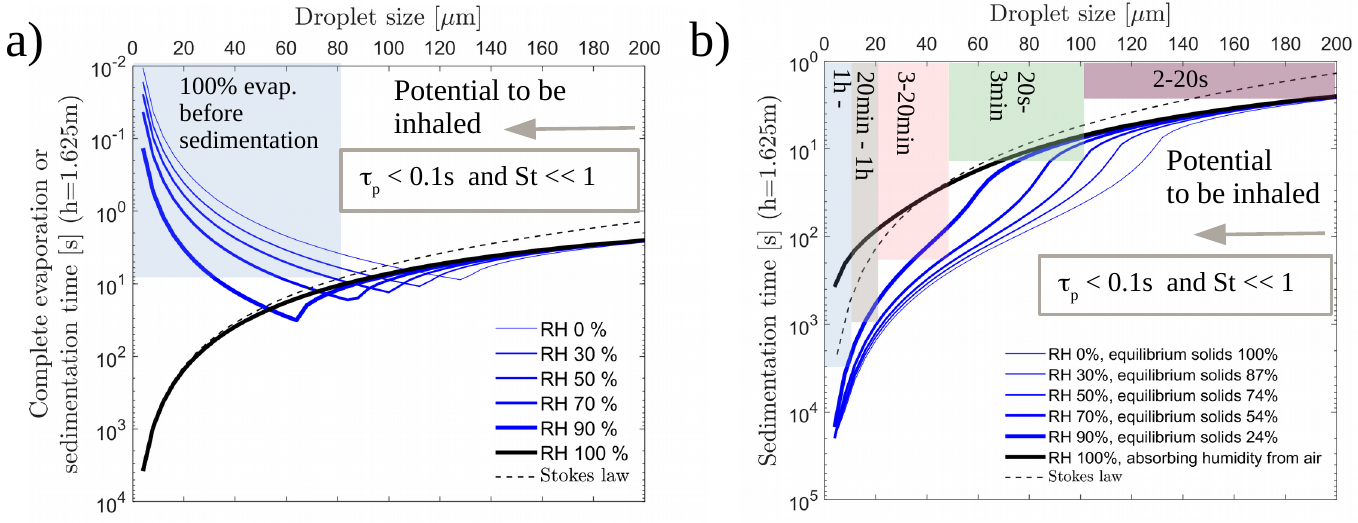}
	}
	\caption{1D simulations on evaporating droplets assuming stagnant ambient air. a) Evaporation time, i.e. Wells curve, for water. b) Sedimentation time of water droplets with non-volatile content. The shaded areas indicate the sedimentation time range. Clearly, all droplets in the considered range of below $200\mu$ m could stay airborne, and hence to be inhaled, as $St\ll 1$ and $\tau_p \ll t_s $. This is particularly true for droplets below 100$\mu$m.}
	 \label{fig:image19}.
\end{figure}

Noting the well-known $t_{evap} \propto d^2$ law for droplet evaporation time, along with $t_s \propto d^{-2}$ for settling, one anticipates a separation between small and large droplets i.e.~droplets that dry quickly before reaching ground and linger in the air, and droplets that settle to ground before the evaporation is complete. In order to assess the matter, we reproduce the classical Wells' evaporation-falling curve from the 1930's using a model similar to that of \citet{Xie2007}. We compare (i) pure water with (ii) water with 3~w-\% non-volatile matter with average molar mass of $50~\unit{g\,mol^{-1}}$. Ideal liquid (Raoult’s law) is assumed in the latter analysis. The model incorporates the time variation of droplet temperature along with the dynamic variation of the droplet radius due to evaporation (RH $\leq$ 100\%) or absorption of moisture in case of high relative humidity in the presence of non-volatile matter. Focusing on typical indoor conditions RH $\leq$ 50\%, Fig.~\ref{fig:image19} a) shows that the evaporation time for water droplets of size $d<80~\unit{\mu m}$ is below their sedimentation time. Hence, they vanish before reaching the ground while the larger droplets $d\leq 80~\unit{\mu m}$ reach the ground before the evaporation is complete. Of course, assuming the ambient flow is stagnant. Fig.~\ref{fig:image19} b) shows that with non-volatile content, all droplets below $50~\unit{\mu m}$ will stay airborne for at least 3 minutes. The final equilibrium moisture in the produced aerosols depends on RH. This may have an effect on the viability of the pathogens and is an interesting additional result of modeling evaporating droplets with non-volatile content. 

As expected, the classical Stokes law predicts sedimentation of the smallest droplets for pure water at $\mbox{RH}=100\%$ when their size does not change. If non-volatile matter is present, very small droplets absorb humidity from air and fall to the ground slightly faster than pure water droplets, but only if the air is saturated or very close to such conditions ($\mbox{RH}=100\%$). A further study was also carried out by assuming evaporation suppression due to surfactants (discussed in Section~\ref{sec:surfactants}). Based on our background numerical work (not shown herein), the presence of surfactants may suppress evaporation to some extent, but the main conclusions presented herein remain unaffected. 

The present analysis indicates that droplets typical for coughing and speaking ($d<20\mu\unit{m}$) are airborne either immediately or via rapid drying in typical indoor conditions ( e.g. $\mbox{RH}<50\%$). We also note that our estimates on evaporation timescales are consistent with the values reported previously by others \citep{Lenhart2004, Xie2007, Stadnytskyi2020}. 

\begin{figure}[H]
	\centering
	\makebox[0pt]{
		\includegraphics[width=1.0\linewidth]{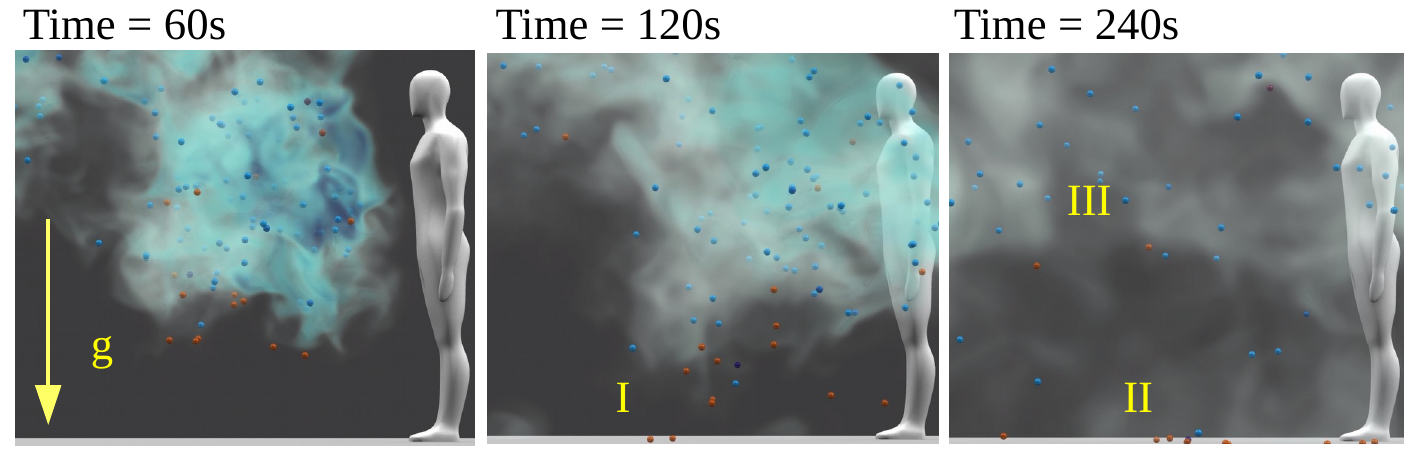}
	}
	\caption{Illustration of particle sedimentation in turbulent flow with RMS velocity $0.02~\unit{m\,s^{-1}}$ based on LES with the NS3dLab code. If these particles were droplets, their drying timescale into droplet nuclei would be so short that even the red particles ($10~\unit{\mu m}< d\leq 20~\unit{\mu m}$) would remain suspended in the air similar to the blue particles ($d\leq 10~\unit{\mu m}$) or smoke.}
	 \label{fig:NS3dLab}.
\end{figure}

\subsection{On the role of airflow and turbulence}

For a typical turbulent air flow, particles follow the flow if their Stokes number $St=\tau_p/\tau_f \ll 1$ i.e. $\tau_p \ll \tau_f$ where $\tau_f$ is the fluid timescale being commonly $O(1~\unit{s})-O(100~\unit{s})$ in indoor environment. If $St \sim  1$, particles may exhibit preferential accumulation with high local concentrations while they hardly interact with the ambient airflow if $St \gg 1$ \citep{Elghobashi}. We focus on droplets with $St \ll 1$ and Weber number $We \ll 1$ i.e. spherical droplets. Fig.~\ref{fig:NS3dLab} illustrates the role of $t_s$ on the sedimentation of solid particles in a 3D turbulent flow at a moderate ambient turbulence level. If the particles would be replaced by evaporating droplets, the largest droplets, with initial $\tau_{evap}\gg t_s$, would reach the ground within minutes (I-II). If $\tau_{evap}\approx t_s$, the droplets may have a chance to become droplet nuclei and start following the flow from that point on since $St\ll 1$. For $\tau_{evap}\ll t_s$ the droplets would dry immediately and follow the flow from that point on. Clearly, most of the sub $10~\unit{\mu m}$ solid particles remain suspended in the air, similar to smoke, during the time of observation (III). 


\subsection{On the role of surfactants}
\label{sec:surfactants}

Viral droplets can be generated in the airways via surface shear and film rupture \citep{Thomas2012}, particularly, via small airway reopening \citep{Bake2019}. Therefore, the droplets contain surfactants and mucus from the respiratory tract. Both surfactants and mucus may affect the droplet drying process, hence, they may also have an impact on the aerosolization of the larger droplets. On one hand, the lifetime of the viruses may be prolonged due to surfactants released during coughing. On the other hand, the surfactants, which are enriched at the air–liquid interface of the respiratory fluid droplets, were observed by \citet{Vejerano2018} to slow down evaporation by factor 2-10, depending on RH. The previous research has focused on liquid–air interfaces that are flat or have less surface curvature than the actual virus-containing airborne droplets \citep{Yu2016,Hermans2015,Zuo2006}. Different compositions of lung surfactants involved with relevant curved surfaces have a distinct interfacial rheology \citep{Hermans2015} that alters the drainage \citep{Bhamla2014}. Along with lung surfactants, mucus within the droplets affects the air-water interfacial rheology \citep{Georgiades2014,Vasudevan2007,Thomas2012}, and may therefore influence the duration of the droplets in air. Mucus has, furthermore, a particular significance for virulence, since the proteins and surfactants within mucus can protect the viruses inside a droplet, prolonging their viability, even after the water has mostly evaporated \citep{Vejerano2018}.


\section{Models, methods and software}
\label{sec:models}

The case setups and the source codes behind the presented numerical simulations will be made openly available during 2020. Hence, the baseline simulations can be reproduced and further modified using the shared benchmarks.  

\subsection{Computational fluid dynamics modelling}

In the present consortium, four different open source CFD software were used: 1) PALM, 2) OpenFOAM, 3) NS3dLab, and 4) Fire Dynamics Simulator (FDS). While 1-3 were used to assess the dilution rate, the subsequent analysis primarily relies on results obtained from PALM simulations. The FDS solver is employed to investigate the effect of air circulating ventilation. In each CFD solver, the Navier-Stokes equations are solved for the velocity field and the evolution of a cough-released aerosol cloud is modelled by solving an additional transport equation for a scalar concentration field or employing a discrete Lagrangian stochastic particle model.    

PALM is a LES solver designed for atmospheric boundary-layer simulations but adapted here for indoor simulations. The code solves the incompressible Navier-Stokes equations in Boussinesq-approximated form using 3rd-order Runge-Kutta time integration \citep{Williamson1980}. Spatial discretization is based on the finite-difference approach with the 5th-order accurate upwind biased advection scheme \citep{Wicker2002} on structural staggered grids. Sub-grid scale turbulence is modelled using a 1.5-order closure based on \citet{Deardorff1980}. PALM is highly optimized for high performance computing environments exhibiting excellent scalability on massively parallelized simulations. More information on PALM is found in \citet{Maronga2015,Maronga2020}.

OpenFOAM is a numerical library employing the finite volume method with 2nd order spatial and temporal accuracy. In this study the buoyantPimpleFoam solver \citep{Weller1998} (version 7) is used in which the continuity, momentum and energy equations are solved in the low Mach number formulation. Convection terms are discretized with a Gamma-flux limited scheme \citep{Jasak1999} while linear schemes are used for all the other terms. We utilize the implicit LES approach as a stand-alone subgrid model similar to our previous studies \citep{Peltonen2018,Peltonen2019,Laurila2019}. 

NS3dLab is a pseudo-spectral solver based on the DNSLab packege implemented in the Matlab language \citep{Vuorinen2016}. Here, the solver is used to assess mixing of a passive scalar and sedimentation of droplets in a periodic, homogeneous turbulence configuration. NS3dLab relies on the fourth order Runge-Kutta time discretization, skew-symmetric (kinetic energy conserving) form of the convection terms, and sixth order explicit filtering of flow variables as a LES model. Further specifications on the solver are found in \citet{Vuorinen2016}.  

FDS is a LES solver for buoyancy driven low-Mach number flows, using structured, uniform and staggered grids, explicit, second-order, kinetic-energy-conserving numerics and simple immersed boundary method for flow obstructions. In this work, we used FDS version 6.7.3 (FDS6.7.4-0-gbfaa110-release) with Deardorff turbulence model and WALE model near the walls. For the aerosol simulation, we used an Eulerian aerosol tracking with near-wall deposition mechanisms. More information can be found in \citet{McGrattan2019,McGrattan2012}.

\subsection{Computational setup}

\begin{figure}[H]
	\centering
		\includegraphics[width=\linewidth]{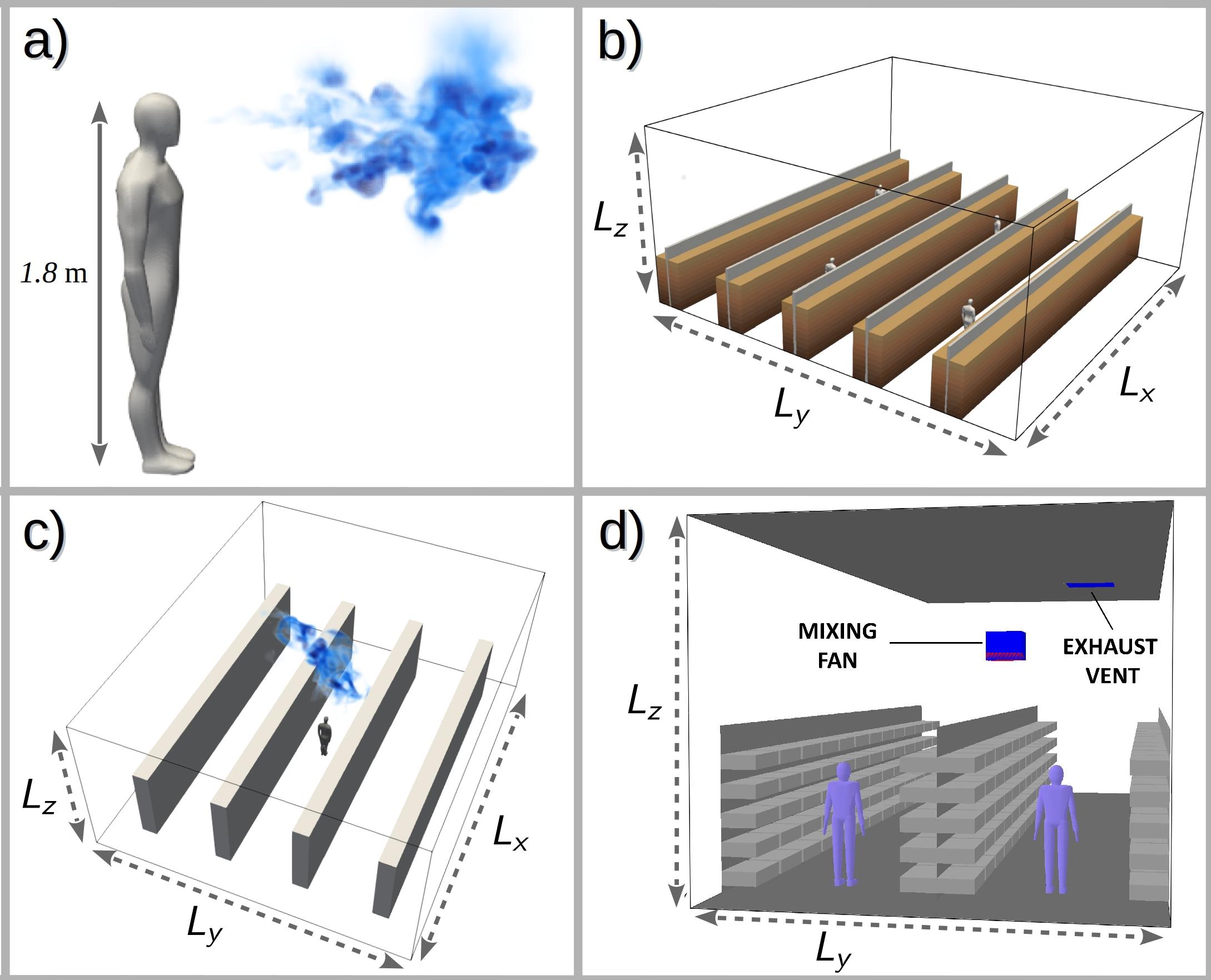}
	\caption{a) Illustration of a representative coughing person, visualizing aerosol cloud concentration, turbulence and mixing. Overviews of the computational models utilized in b) PALM, c) OpenFOAM and d) FDS simulations.}
	\label{fig:computational_setup}
\end{figure}

Illustrations of the studied setup are shown in Fig.~\ref{fig:computational_setup} along with the main dimensions listed in Table~\ref{tab:setup_dims_gridsize}. While the general system dimensions were fixed, the case setups differed to some extent in terms of the boundary conditions, and the turbulence generation procedure. 

In the PALM simulations (Fig.~\ref{fig:computational_setup} b), model shelves are comprised of $2.5~\unit{m}$ high solid dividers separating $2~\unit{m}$ high porous media blocks.
Using porosity allows flow penetration into the shelf-space and qualitatively mimics the drag caused by the shelf structures and the goods on them. The lateral boundary conditions are periodic for the flow field while zero condition was given to the passive scalar. Particles hitting the side boundaries were simply removed.

In OpenFOAM (Fig.~\ref{fig:computational_setup} c) four $2.5~\unit{m}$ height shelves are modelled. The shelves are modelled as impermeable rectangular blocks and the aisle width is $2~\unit{m}$. Single cough source is placed in the middle of the domain. Periodicity is applied at the side boundaries for all resolved fields while the no-slip condition is applied at the floor, ceiling, shelves and the coughing person. 

The FDS setup (Fig.~\ref{fig:computational_setup} d) is perhaps the most exploratory what comes to the shelf-assembly modelling as each shelf plate is separately modeled. In the FDS setup, two $13~\unit{m}$ long and $2~\unit{m}$ wide aisles were considered. A coughing person is placed on each of the aisles. The aisles were surrounded by $0.5~\unit{m}$ deep and $0.2~\unit{m}$ high obstructions with $0.2~\unit{m}$ separations modelling the shelves. The total height of the shelves is $2.5~\unit{m}$. The side boundaries in the FDS setup are periodic for the flow field but open for the aerosol-tracers. 

\begin{table}[!t]
	\centering
	\caption{Summary of computational model details. Here, PALM, OpenFOAM and NS3dLab are used to estimate dilution time and FDS is used to assess ventilation aspects.}
	\label{tab:setup_dims_gridsize}
	\includegraphics[width=\linewidth]{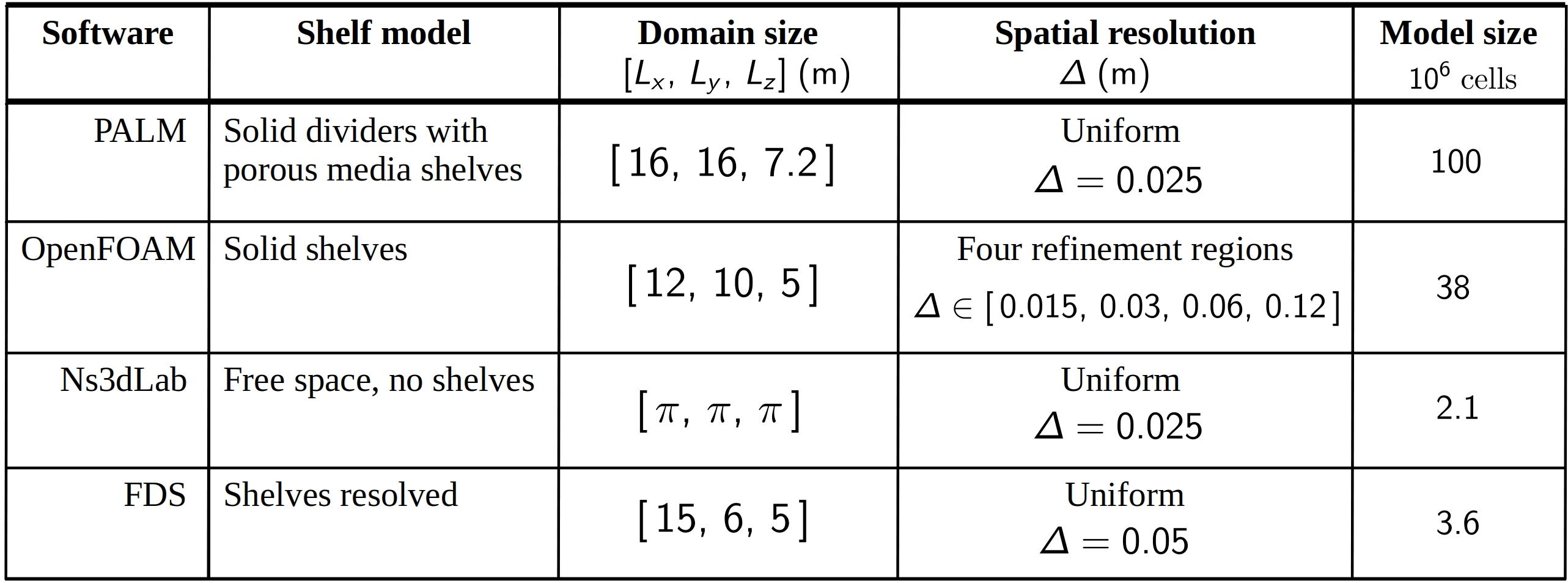}
\end{table}

\subsubsection{Background flow}

The subsequent description on modelling approaches is intentionally brief. For more detailed information on the CFD simulations, refer to the case setup materials which are made openly available to ensure transparency and reproducibility.

In PALM and OpenFOAM simulations, the turbulent flow field within the retail-store domain are generated by applying a source term to the momentum equations. The implementations vary such that PALM model features a specific ventilation flow zone near the roof (for $z > 0.7L_z$) where the flow is locally driven by a constant force, energizing the turbulence below while maintaining a low mean flow level. The OpenFOAM setup utilizes a time-varying perturbation force to first initialize the ambient turbulence while, prior to the cough event, switching on a constant body force to impose a specific mean speed and direction. The initial development of the ambient turbulence field is monitored and the cough event is initiated only after the turbulence conditions have settled. 

Two mean flow directions are studied in PALM and OpenFOAM simulations: $\alpha=10^\circ$ and $\alpha=90^\circ$ w.r.t. the direction of the aisle. The angle $\alpha=0^\circ$ indicates a direction along the aisle. Through experiments, in the context of this study, these two flow directions were determined to capture the relevant range of flow phenomena.   

In FDS simulations, the flow is driven by a local ventilation machine which circulates air. The device is situated above the first aisle (right hand -side aisle in Fig.~\ref{fig:computational_setup} d) at $3~\unit{m}$ height. The air is sucked into the device from above and released downward via two $0.1~\unit{m^3s^{-1}}$ air streams in $\theta=45^\circ$ angle from vertical, one along the aisle in positive $x$-direction, and another in negative $x$-direction. The resulting air speeds in the first aisle were in the range $2-20~\unit{cm\,s^{-1}}$ and in the second aisle $1-5~\unit{cm\,s^{-1}}$. In addition, exhaust vents were added in some simulations to the ceiling as a $1~\unit{m^2}$ square above the first aisle. The characteristic feature of these simulations is the downwards air stream in the region of the coughing person on the first aisle.

\subsubsection{Model of a coughing person}

In all featured simulations, the coughing human is described in a simplified manner as a $1.8~\unit{m}$ tall human-like figure or a solid block. The cough is released at assumed mouth height $1.6~\unit{m}$. (The simulation results are not sensitive to the spatial details of the coughing person.) A more detailed human model is used in the subsequent illustrations for improved visual effect. 

A cough or a sneeze is associated with a rapid burst of exhaled air where all the mass and momentum is released into the surroundings within $\approx 0.3$ s which is a characteristic cough duration \citep{Lindsley2013}. Because the spreading and dissipation of the resulting cough plume occurs over a much longer time span by the surrounding turbulence, the actual resolution of the initial cough event is not critically important in the considered simulations. However, it is essential that modelling of the cough phenomena correctly accounts for resulting momentum and energy input into the surrounding space. 

In this work, PALM and OpenFOAM simulations employed compatible approaches in initializing the cough plume. This involved releasing $V_{\mathrm{cough}} = 1.375~\unit{dm^3}$ at temperature $36~\unit{^\circ C}$ within the cough duration $T_{\mathrm{cough}} = 0.3~\unit{s}$ through a mouth area $A_m = 9\times 10^{-4}~\unit{m^2}$ \citep{Bourouiba2014}. The ambient air temperature is $20~\unit{^\circ C}$. The OpenFOAM simulations exploited the local refinement capability and incorporated the initial cough event into the retail-store simulations whereas the PALM simulations employed an integral approach, which utilized the result from an isolated, very high-resolution OpenFOAM simulation on the initial cough plume development, adopting the mean state and shape of the momentum and energy plume at $t=0.3~\unit{s}$. (During this initial phase the ambient conditions have not been able to affect the plume development.) 

In the FDS simulations, the cough volume was set to $V_{\mathrm{cough}} = 4.2~\unit{dm^3}$, released within $ 0.3~\unit{s}$ \citep{Lindsley2013} from a rectangular surface of $A_m = 9\times 10^{-4}~\unit{m^2}$. The cough had a water mass fraction of 3\% and temperature $32~\unit{^\circ C}$. 

In all simulations the spreading and dissipation of the aerosol-laden air is modelled via Eulerian approach, where a fixed concentration is assigned to the cough volume, and the subsequent evolution of the concentration is computed by solving a scalar conservation equation. In the PALM simulations, Lagrangian stochastic particle model with $6 \times 10^6$ particles is employed alongside with this scalar approach, allowing the particle size effect to be studied as well. The large number of released particles is necessary to obtain a sufficiently accurate description of the cough plume dispersion which remains comparable to the scalar approach. In the NS3dLab simulations a small number of particles is studied. 

\subsection{Monte-Carlo modelling}
\begin{figure}[H]
\centering
  \includegraphics[width=0.95\linewidth]{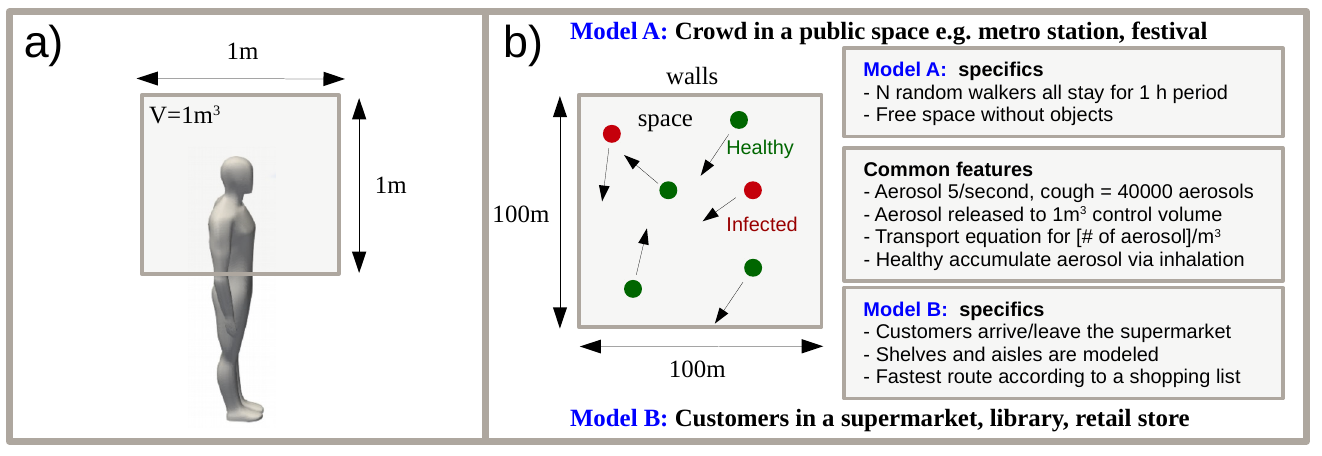}
\caption{a) Both Monte-Carlo models assume release of aerosol into a control volume $V$ by the infected individuals. Dilution is incorporated as diffusion and removal processes. b) Models A and B rely on the same core assumptions. Model A assumes a generic space while model B targets a generic supermarket configuration.}
\label{fig:MCsetup}
\end{figure}

As summarized in Fig.~\ref{fig:MCsetup}, we develop two Monte-Carlo models in order to understand a transmission scenario via inhalation of aerosols. Two fast tools, relying on the same model equations, are implemented to incorporate the spatio-temporal aerosol dispersion. The tools were required to 1) have the capability of tracking individuals as an event-based model i.e. follow the paths of individual persons \citep{Helbing1995,Bonabeau2002,Helbing2005,Kim2013}, 2) enable input from the CFD-simulations regarding the properties of the plumes and various aerosol production possibilities, 3) allow control over basic metrics such as number of persons, coughing intensity, percentage of infected individuals, and 4) be computational efficient. To our understanding, the model described herein has not been developed earlier by others.

In model A, implemented in the Matlab language, we model $N$ individuals in a 2D square area of size $A=100~\unit{m} \times 100~\unit{m}$ walking at speed $U$ from a uniformly distributed random position to another for a fixed duration of 1h. The infected persons amount to $qN$ with $q=0.005,\,0.01,\,0.02$ representing three possible scenarios for the second phase of COVID-19 ($0.5-2\%$ of the population). Depending on the fixed average number density of persons ($N/A$), model A could represent any generic public place such as a mall (e.g. $N/A \leq 0.02~\unit{m^{-2}}$), metro station (e.g. $N/A \leq 0.5~\unit{m^{-2}}$), or public transportation/bar (e.g. $N/A\leq 1~\unit{m^{-2}}$). Also the walking speed of the individuals is discussed in the paper. In model B, implemented in the Python language, a generic supermarket environment, with entrance, exit and cashiers, is modeled along with aisles surrounded by shelves. The relevant $N/A < 0.02~\unit{m^{-2}}$ on average. In contrast to model A, the number of customers is not fixed but they arrive to the system at random times and leave once they have completed their randomized shopping list following an adaptation of a shortest path algorithm. 

In both models, we solve a diffusion equation for the number density of aerosols $c=c(x,y,t)$ ([$c$]=$\unit{m^{-3}}$) assuming that the infected persons release aerosol at a constant rate $\lambda = 5~\unit{s^{-1}}$ \citep{Asadi2019}. Two scenarios, with and without coughing, are modeled at the occurrence rates of 6 or 0 times an hour i.e. $p_c=6/3600~\unit{s}$ and $p_c=0$, respectively. During a single cough, we assume instant release of $N_c = 40,000$ aerosol particles at the location of the person with relevance to the study by \citet{Lindsley2012}. Hence, the expectation value of the aerosol generation rate due to coughing is $\lambda_c=p_c N_c$. We assume that all the released aerosol is immediately mixed to a control volume of $V=1\unit{m^3}$ after which the aerosol starts to diffuse in $x,y-$directions. The diffusion equation
\begin{equation} \label{eq:diff}
\frac{\partial{c}}{\partial{t}} = D\Delta c + S - c/\tau
\end{equation}   
represents the spatial mixing of the aerosol in the air due to the air motion.
Here, the aerosol source term $S=\Sigma_{k}S_k(x-x_k,y-y_k,t)$ with $\{x_k,y_k\}_{k=1}^{qN}$ being the transient location of the infected persons. Removal of aerosol by e.g. ventilation or vertical surfaces is modeled by a sink term $-c/\tau$ with a removal timescale of $\tau=100~\unit{s}$ with relevance to the dilution timescale obtained from the CFD simulations. Aerosol is produced locally at the random positions of the walkers. The diffusion constant $D=0.05\unit{m^2/s}$. 

Eq.~\ref{eq:diff} is solved by a standard second order finite difference model with explicit Euler time stepping. Within each time step, the diffusion equation is solved using the source terms evaluated using deterministic and stochastic components. Within $t...t+\Delta t$, an infected person releases aerosol $\lambda \Delta t$ units. In contrast, coughing occurs during the time step if $p_c\Delta t < r$, where $r\in [0,1]$ is a uniformly distributed random number. In a steady state, the source and sink terms of Eq.~\ref{eq:diff} are in balance and we see that the average aerosol concentration in the whole system ($[c_{ave}]=\unit{1/m^3}$) is: 
\begin{equation} \label{eq:cave}
c_{ave} = \frac{qN\tau}{A h}(\lambda + \lambda_{c}), 
\end{equation}
where $A$ is a representative area of the system, such as the floor area (here: $A_f=10,000~\unit{m^2}$), and $h$ is the control volume height (here: 1$\unit{m}$). For example, the value $c_{ave}=1000~\unit{m^{-3}}$ would indicate exposure to 1200 aerosol particles during a one hour inhalation period for a typical consumption of $1.2~\unit{m^3}$ per hour. Model functionality can be verified using Eqs.~\ref{eq:cave} and \ref{eq:pave}, see Section~\ref{sec:results}. The simulated individuals accumulate aerosol proportional to their rate of inhalation for which we assume the average rate $\dot{V}_b=0.33~\unit{dm^3\,s^{-1}}$. Hence, the average number of aerosols a person inhales, increases linearly in time according to the following exposure law
\begin{equation} \label{eq:pave}
N_b(t) = c_{ave} \dot{V}_b t. 
\end{equation}   
In fact, Eq.~\ref{eq:pave} turns out to be predictive for the average exposure obtained in the Monte-Carlo model. We note that in practice $N/A$ is seldom a constant as people e.g. visit more probably certain compartments of a retail store, rush in to the school via same corridors at the same time etc. For example, in Model A, despite the target points being uniformly distributed, the probability of finding the individuals within radius $L/2 = 50~\unit{m}$ from the domain center is approximately 95\% i.e. the actual individual positions are not uniformly distributed. Hence, we use in practice a more representative floor area of $A=\frac{\pi}{4}A_f$ for model verification \footnote{$A_{circle}/A_{square}=\frac{\pi}{4}\approx 0.785<0.95$}. 

\section{Results and discussion}
\label{sec:results}

\subsection{Background}
The present modelling work focuses on pathogen transmission as aerosols or small droplets. The results of the paper represent generic model scenarios focusing on transmission by inhalation. The presented results from the analysis could in many circumstances be re-scaled to incorporate e.g. up-to-date coronavirus-related information when available. The results could also be re-scaled to assess transmission of other viruses or to account for different amounts of cough released aerosols.  

Our main results consist of 1) order of magnitude estimates on the number of aerosol particles that the infected persons from publicly known case studies have been able to inhale (yet, we do not know did they actually get the infection this way), 2) 3D CFD (LES) based numerical demonstration on the air flow and aerosol cloud physics in a generic public place such as an open office space or a supermarket, 3) numerical estimates on the cough plume dilution timescale along with the critical exposure times and distances to cough plumes, 4) Monte-Carlo-based numerical assessment on transmission scenarios in generic public places, focusing only on transmission by inhalation, and 5) CFD based demonstration on the coupling between a ventilation system and an aerosol cloud.

\begin{table}[!t]
	\centering
	\caption{Reported airborne transmission cases from the media. HS stands for Helsingin Sanomat, the largest subscription newspaper in Finland, and YLE stands for Finland's national public broadcasting company Yleisradio.}
	\label{tab:tab3}
	\begin{tabular}{p{1.7cm}p{5cm}p{3.3cm}p{2.3 cm}}
    	\hline
    	Reference & Case & Present assumptions & Estimated number of inhaled aerosols $N$\\ \hline


        HS 1 & Sipoo wedding Mar 7, 2020: 96 persons, ‘several’ were infected, duration few hours, no symptomatic persons. 
                &  \vspace{-1.5ex} 
                   \begin{tabular}{l}
                    $V = 1600~\unit{m^3}$ \\ $\lambda = 1 - 50~\unit{s^{-1}}$ \\ $t~=~3~\unit{h}$  
                    \end{tabular} 
                        & $N\sim 50 - 500$ \\[1em]
        HS 2 & Sipoo birthday party Mar 7, 2020. 70 guests, more than 10\% got infection, no symptomatic persons. 
                &  \vspace{-1.5ex}
                   \begin{tabular}{l}
                    $ V  = 800~\unit{m^3}$ \\ $\lambda = 1 - 50~\unit{s^{-1}}$ \\ $t~=~3~\unit{h}$  
                   \end{tabular}   
                        & $N\sim~100~-~1000$ \\[0.5em]
        \citet{Lu2020} & Restaurant, Guangzhou, China, Jan 24: 91 persons ate at the restaurant, 10 infected. Four infected persons sat in the same table as the asymptomatic index patient, three were in contact for 53 minutes and two for 73 minutes. Dining area was $145~\unit{m^2}$. 
                &  \vspace{-1.5ex}
                   \begin{tabular}{l} 
                     $V = 400~\unit{m^3}$ \\ $\lambda = 1 - 50~\unit{s^{-1}}$ \\ $t = 1~\unit{h}$  
                   \end{tabular}  
                        & $N\sim 20 - 200$ \\[0.5em]
        HS/YLE 3 & Women's Day concert, Mar 8, 2020, Helsinki. $< 1700$ persons, at least 7 infected, duration $2~\unit{h}$. 
                &  \vspace{-1.5ex}
                    \begin{tabular}{l} 
                    $V = 10,000~\unit{m^3}$ \\ $\lambda =\mbox{log-normal}$  \\ $t = 2~\unit{h}$
                    \end{tabular}  
                        &  $N\sim 30 - 400$ \\
        \hline
\end{tabular}
\end{table}

\subsection{Estimates on the critical level of aerosol exposure}
\label{Sec:critical_exposure}

While the critical dose of the virus-containing particles is not known, order of magnitude estimates on the aerosol exposure can be deduced from incidents reported during the ongoing COVID-19 pandemic. Not all aerosols would carry viruses but part of them could. Table~\ref{tab:tab3} lists a few of such incidents where, according to the media reports, several people were infected. 

The coupling between the particle production and the ventilation rate could be modelled using classical methods, such as the Wells-Riley approach~\citep{Riley1978}. Here, we use a simplified model assuming a single individual emitting aerosol particles at rate $\lambda$. The particles are instantaneously mixing in a space with volume $V$, very low ventilation exchange rate, and no initial particles. Under such circumstances, the mean particle concentration increases steadily during the gathering, and we can calculate the number of inhaled particles for each person as 
\begin{equation}
 N_b(t) = \int_0^t \dot{V}_b c(t') dt'
 =\dot{V}_b \int_0^t \frac{\lambda} {V} t' dt' = \frac{\dot{V}_b \lambda t^2} {2V}.
\end{equation}

For cases with a pre-symptomatic emitter, we assume $\lambda = 1-50~\unit{s^{-1}}$ \citep{Asadi2019}, and for coughing, $\lambda $ was calculated as discrete releases of $N$ particles every 5~min, where $N$ was picked randomly from a log-normal distribution with parameters $(10.74,0.99)$ \citep{Lindsley2012}. The breathing rate $\dot{V}_b$ was assumed to be a uniformly distributed random variable between $8~\unit{dm^3\,min^{-1}}$ and $20~\unit{dm^3\,min^{-1}}$. Assigning estimated space volume and event duration and using the 10\% and 90\% fractiles as the lower and upper bounds, we get the range $10\leq N_b \leq 1000$ for the number of inhaled particles in the considered four cases. In the following analyses, we assume a critical number of inhaled particles $N_{b,\mathrm{cr}}= 100$. The above assumptions should be refined if official investigations become available, and the inherent uncertainty of this estimate should be kept in mind when evaluating the results.

\subsection{Flow visualization of exhaled air and droplet size effects}

\begin{figure}[H]
	\centering
	\makebox[0pt]{
		\includegraphics[width=0.9\linewidth]{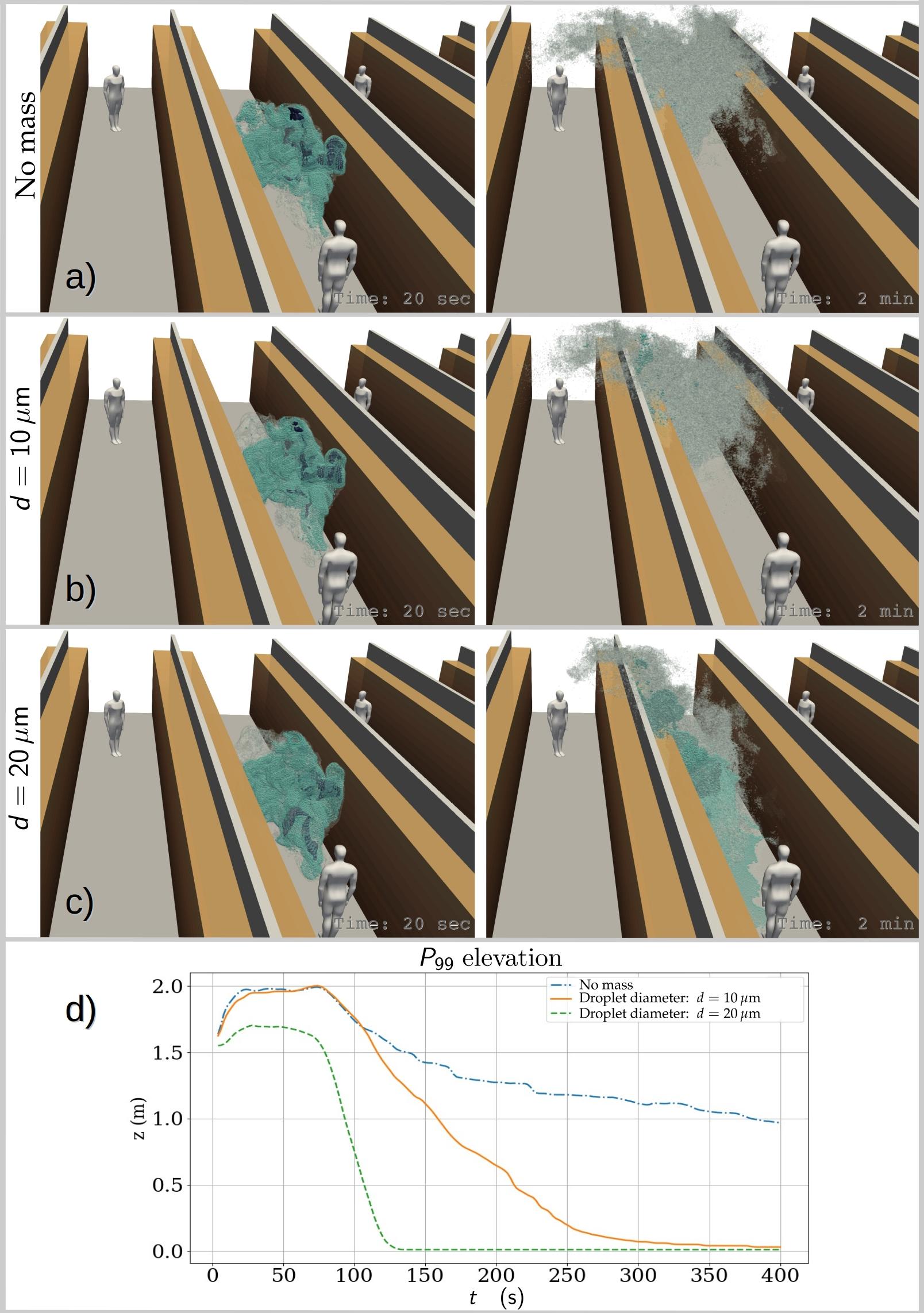}
	}
	\caption{Visualizations demonstrating the effect of particle size (and mass) on the modelled spreading of the cough-released aerosol cloud. For better sense of scale, bystanders are placed 8~m from the coughing person. Instantaneous views on the state of the cloud are shown for realizations where the particles have a) no mass, b) $1000~\unit{kg\,m^{-3}}$ density and $10~\unit{\mu m}$ diameter and c) $1000~\unit{kg\,m^{-3}}$ density and $20~\unit{\mu m}$ diameter. Images on the left column are at $t=20~\unit{s}$ and on the right column at $t=120~\unit{s}$. Below, d) presents the time evolution of the mean elevation of the 99th percentile concentration highlighting the different descent rates. Droplets in these size scales have $\tau_{evap}<1s$ and they would become aerosol-like droplet nuclei very rapidly. }
	\label{fig:image14}
\end{figure}

To give an overall impression on the dispersion evolution of the cough-released aerosol cloud, Fig.~\ref{fig:image14} depicts concentration isosurfaces at two different time instances obtained from PALM simulations where the Lagrangian particles have been assigned a) no mass, b) $10~\unit{\mu m}$ diameter and c) $20~\unit{\mu m}$ diameter with $1000~\unit{kg\,m^{-3}}$ density. The left column images are at $t = 20~\unit{s}$, which is relatively early in the cloud development, while the right column images exhibit the state at $t = 120~\unit{s}$. Gravity and excess inertia play very weak role in the dynamics of particles smaller than about $10~\unit{\mu m}$ at the present time scales. Therefore it is expected that the simplifying assumption of massless particles is justified when modelling particles smaller than about $10~\unit{\mu m}$ as $St\ll 1$.

The massless and $10~\unit{\mu m}$ particles are shown to behave in a closely comparable manner while a noticeable part of the $20~\unit{\mu m}$ particles show remarkably different behaviour especially at later times. Fig.~\ref{fig:image14}~d) shows the time evolution of the mean elevation of the 99th percentile concentration highlighting the different descent rates. This plot reveals that also the $10~\unit{\mu m}$ particles do descent to some extent at later times. However, closer inspection revealed that these $P_{99}$ elevation curves exaggerate the differences, especially those between massless and $10~\unit{\mu m}$ particles for $t > 100~\unit{s}$. This happens because a small portion of the heavier particles, once they settle on the floor, start to accumulate. As the cloud continues to disperse and dilute, much of the $P_{99}$ is then found on the floor. This shifts the mean elevation rapidly downward. But, a vast majority of the particles behave similarly to massless particles. In this study our focus is mainly on particles smaller than $10~\unit{\mu m}$, hence from now on we assume that massless particle approximation is acceptable in the subsequent analysis. We note that the Stokes number $St\ll 1$ for such particles. This assumption qualifies also application of the passive scalar approach to model the aerosol concentration. The passive scalar approach is employed in all four codes: PALM, OpenFOAM, FDS, and NS3dLab. The Lagrangian particle transport approach is additionally used in PALM and NS3dLab simulations. If the studied particles would have been drying droplets, based on our numerical demonstrations in \ref{sec:droplets}, they would be expected to become droplet nuclei within $\tau_{evap} < 1s$ and follow the massless particle $P_{99}$ curve closely from that time on.

\subsection{Metrics for mixing of exhaled air}

Results from PALM and OpenFOAM simulations featuring two mean flow directions (10 and 90~deg relative to the aisle) are considered herein. As typical to indoor air flows, it is important to note that the simulated flows are characterized by temporally evolving turbulent fluctuations which are of the same order of magnitude as the mean velocity. Therefore, the dispersion realization of any particular cough-released aerosol cloud becomes dependent on the specific arrangement and evolution of the turbulent structures affecting the cloud during its existence. This implies that a single simulated realization may not provide the full picture of this dispersion process. The rigorous way to overcome this problem would be to simulate a large ensemble of realizations and to investigate the ensemble statistics. Unfortunately, this is too time consuming and computationally too heavy a task in this case. Therefore, we decided to carry out a small “mini-ensemble” study consisting of only ten realizations for one set up which is the 10~deg case in order to assess the level of variability due to turbulence. Even though this mini ensemble can not provide us with sufficiently converged statistics, it does offer a more comprehensive view on the studied dispersion phenomena. 

The mini ensemble was obtained by executing 10 consecutive cough simulations, each with identical aerosol cloud initializations, but each released into unique realizations of turbulent flow structures.
This set of simulations took approximately 50 hours using 800 processor cores on the Atos super cluster of CSC – IT Center for Science LTD based on Intel Xeon Cascade Lake processors with 20 cores each running at $2.1~\unit{GHz}$, see \url{https://docs.csc.fi/computing/system/#puhti}. Also all other simulations for this study were run on this system.
 
Next, the predicted dilution curves are studied. For this purpose, dilution curves as functions of both dimensional and nondimensionalized time from various runs are plotted in the same graph. Such runs include the ensemble run and numerous single-realization runs for both 10 deg and 90 deg mean-flow directions, and varied mean-flow rates, including results from the PALM and OpenFOAM simulations. The concept of dilution curve is here defined as the time series of the 99th percentile of the concentration, denoted $P_{99}$, no matter where in the space these highest one-percent values occur. However, the space above the height of 2.2 m is excluded from the analysis. We tested the sensitivity of the dilution curves to the exact choice of the percentile by comparing $P_{99}, P_{98},$ and $P_{96}$ observing  insignificant changes to the quantitative conclusions.  Concentrations are nondimensionalized to unity at the time instance of the cough. Time is nondimensionalized using a turbulent velocity scale $u_{\textsc{tke}}$ and a geometric length scale $L$ such that $t^+ = t \, u_{\textsc{tke}} L^{-1}$. The length scale $L$ is set to the width of the aisle bound by the solid shelves or dividers. This distance is $2~\unit{m}$ in the OpenFOAM setup and $3.15~\unit{m}$ in the PALM setup (because of the porous shelf blocks)

\begin{figure}[H]
	\centering
	\makebox[0pt]{
		\includegraphics[width=\linewidth]{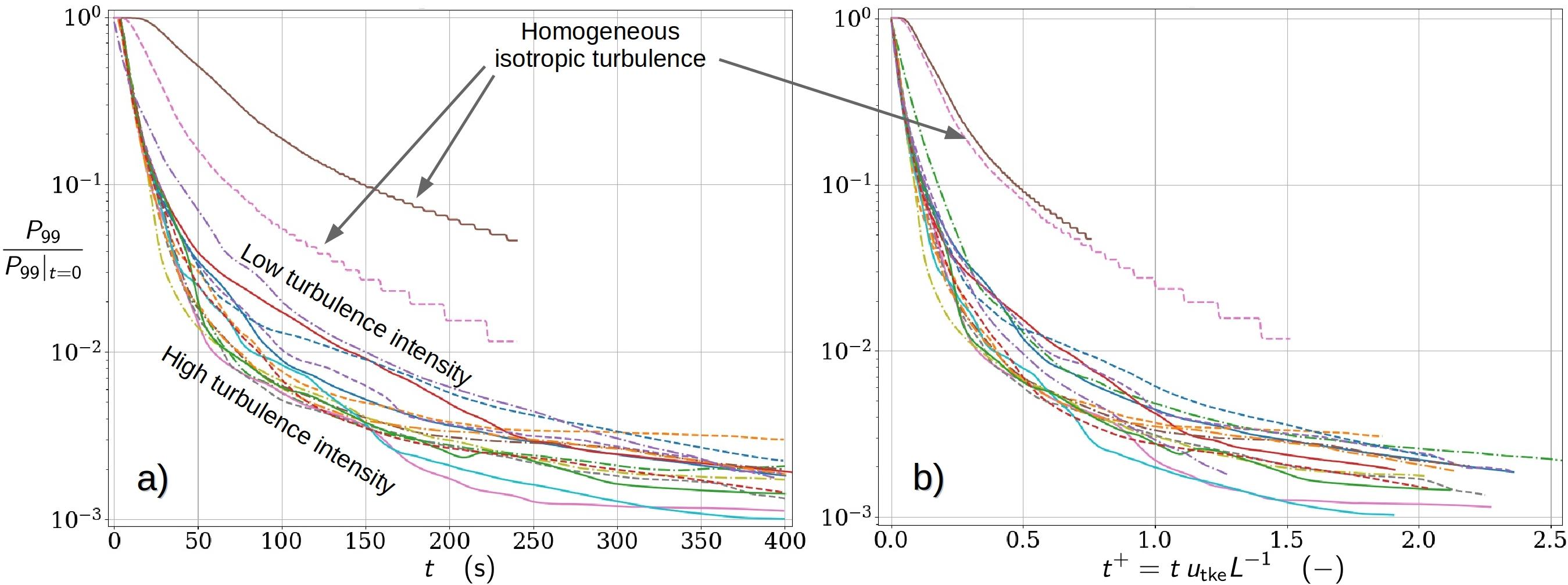}
	}
	\caption{Normalized dilution curves of 99th percentile concentration of massless aerosols plotted against a) time and b) nondimensionalized time $t^{+} = t u_{\textsc{tke}} L^{-1}$ where $u_{\textsc{tke}}$ is the turbulent kinetic energy velocity and $L$ is the characteristic length scale of the flow system. Results are presented for a multitude of simulated cough realizations under varying retail store-like configurations and ventilation conditions. Dilution curves under homogeneous isotropic turbulence conditions using two comparable turbulence intensity levels are also included to highlight the dissimilarity of the studied flow system.}
	\label{fig:image23}
\end{figure}

Fig.~\ref{fig:image23}~a) shows the dilution curves as functions of dimensional time and Fig.~\ref{fig:image23}~b) as functions of nondimensional time. Two of the curves (labeled as homogeneous isotropic turbulence) represent dilution in homogeneous isotropic slowly decaying turbulence with two different initial conditions and they were simulated using the NS3dLab code. In the homogeneous-turbulence simulations the dilution is significantly slower than in the real geometries on average. This difference remains large even after nondimensionalizing the time, but the two homogeneous-turbulence curves collapse to one. This indicates that the dilution depends not only on the turbulence level but also on the geometry and the ambient flow details. We note that the NS3dLab simulations are restricted to a periodic cube ($\pi^3 \unit{m^3}$) in contrast to the very large free space for dilution in the PALM and OpenFOAM simulations. This highlights also the difference between different indoor environments, e.g.~a small office room vs a large lobby. 

How about the significance of the detailed geometrical modelling choices in the dilution analysis? This question is next studied using the set of dilution curves. The collection of dilution curves in Fig.~\ref{fig:image23}~a) from the realistically modelled cases, including the mini ensemble, shows the total variability due to both turbulence and the differences in the set ups (PALM vs OpenFOAM codes and set-ups, mean-flow rates, mean-flow angles). Fig.~\ref{fig:image23}~b) shows the same curves as functions of nondimensional time largely eliminating the variability originating from the differences in indoor configurations. Therefore, the remaining variability is mainly due to the larger turbulence structures that scale with the domain size. Comparison of these plots reveals that the turbulent variability is almost as large as the total variability. However, it is important to bear in mind that the turbulent variability is likely underestimated in Fig.~\ref{fig:image23}~b) due to the limited size of the mini ensemble.

The presented normalized and nondimensionalized dilution curves provide a general description on the relationship between dilution rates and the characteristic turbulence intensity (via $u_{\textsc{tke}}$) within the studied flow system. However, complementary analysis is needed in order to assess the exposure risk. In order to carry out a detailed exposure risk analysis, the problem needs to be dimensionalized again. For this purpose, we choose a representative ambient turbulence-level realization and set 40,000 as the initial number of cough generated aerosols as discussed in Section~\ref{sec:particleDistribution}. Then, the analysis proceeds by 1) re-scaling the initial scalar concentration or 6M Lagrangian particles to represent the value 40,000, and 2) exploiting the outcome of the critical exposure analysis in Section~\ref{Sec:critical_exposure}. In the subsequent quantitative analysis $N_{\mathrm{b,cr}} = 100$ aerosol particles is used as an estimate for the critical exposure. 

\begin{figure}[H]
	\centering
	\makebox[0pt]{
		\includegraphics[width=\linewidth]{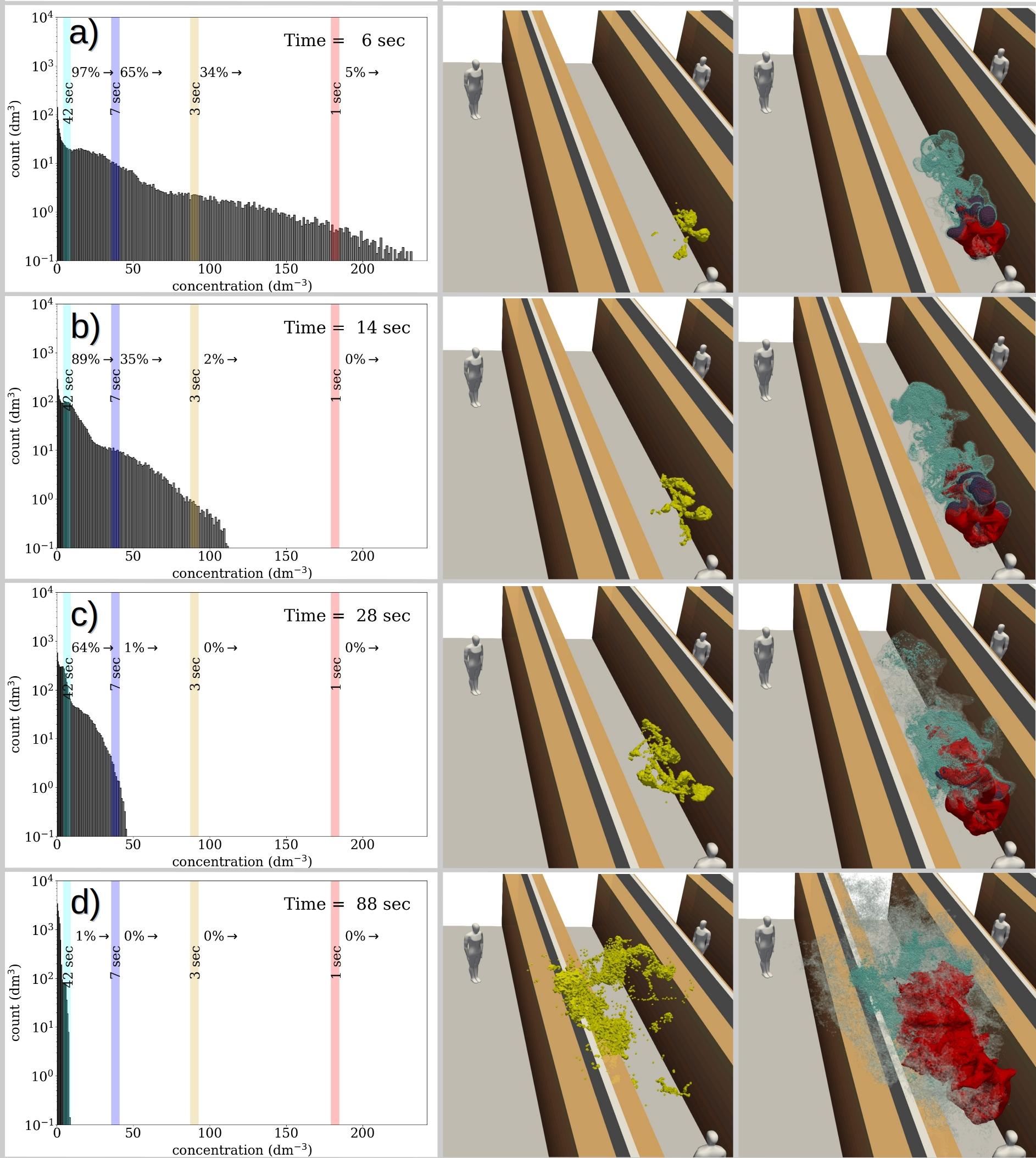}
	}
	\caption{The evolution of the concentration histogram (left column) in juxtaposition with 3-dimensional evolution of the 99th concentration percentile (yellow contour in the center column), and the cloud visualized using three different concentration contours with blue/gray colours (right column). In the right-column images the red contours indicate the DER where a stationary person breathing at rate $0.33~\unit{dm^3\,s^{-1}}$ would have accumulated a critical exposure already. The histogram plots on the left column contain exposure time indicators which denote how long exposure to the marked concentration leads to a critical exposure.}
	\label{fig:image33}
\end{figure}

As a probabilistic approach to the exposure problem, the time-evolution of aerosol concentration distributions via time-evolving concentration histograms are examined. See examples at four different time instances on the left column of Fig.~\ref{fig:image33}. These histograms reveal the dilution behavior of the entire aerosol cloud, detailing the specific volume (in liters) that each concentration level occupies within the expanding (and diluting) cloud. As each concentration level can be paired with a critical exposure time, during which a person with a breathing rate $\dot{V}_b=0.33~\unit{dm^3 \, s^{-1}}$ inhales $N_{\mathrm{{b,cr}}}$ aerosols, the evolving histograms reveal the changing probability distribution for the inhaled aerosol concentrations. For example, the concentration corresponding to $P_{99}$ in the dilution curves, by definition, occur at 1\% probability within the aerosol cloud. 

To illustrate how $P_{99}$ evolves in time, consider the concentration histograms shown on the left column of Fig.~\ref{fig:image33}: during the first a) 7~s ($t^+=0.037$), b) $15~\unit{s}$ ($t^+=0.078$), c) $28~\unit{s}$ ($t^+=0.146$), d) $88~\unit{s}$ ($t^+=0.460$) from the cough release, persons within the cloud are exposed at 1\% probability to $P_{99}$ concentration levels which lead to critical exposure within a) $1~\unit{s}$, b) $3~\unit{s}$, c) $7~\unit{s}$, d) $42~\unit{s}$ or less. Note that in the first two histograms in Fig.~\ref{fig:image33}~a) and b), the probabilities differ from 1\% because of the $2~\unit{s}$ time-resolution of the stored data. In these cases, the above listed dilution times are obtained by interpolating between two time-instants.

\begin{figure}[H]
	\centering
	\makebox[0pt]{
		\includegraphics[width=0.6\linewidth]{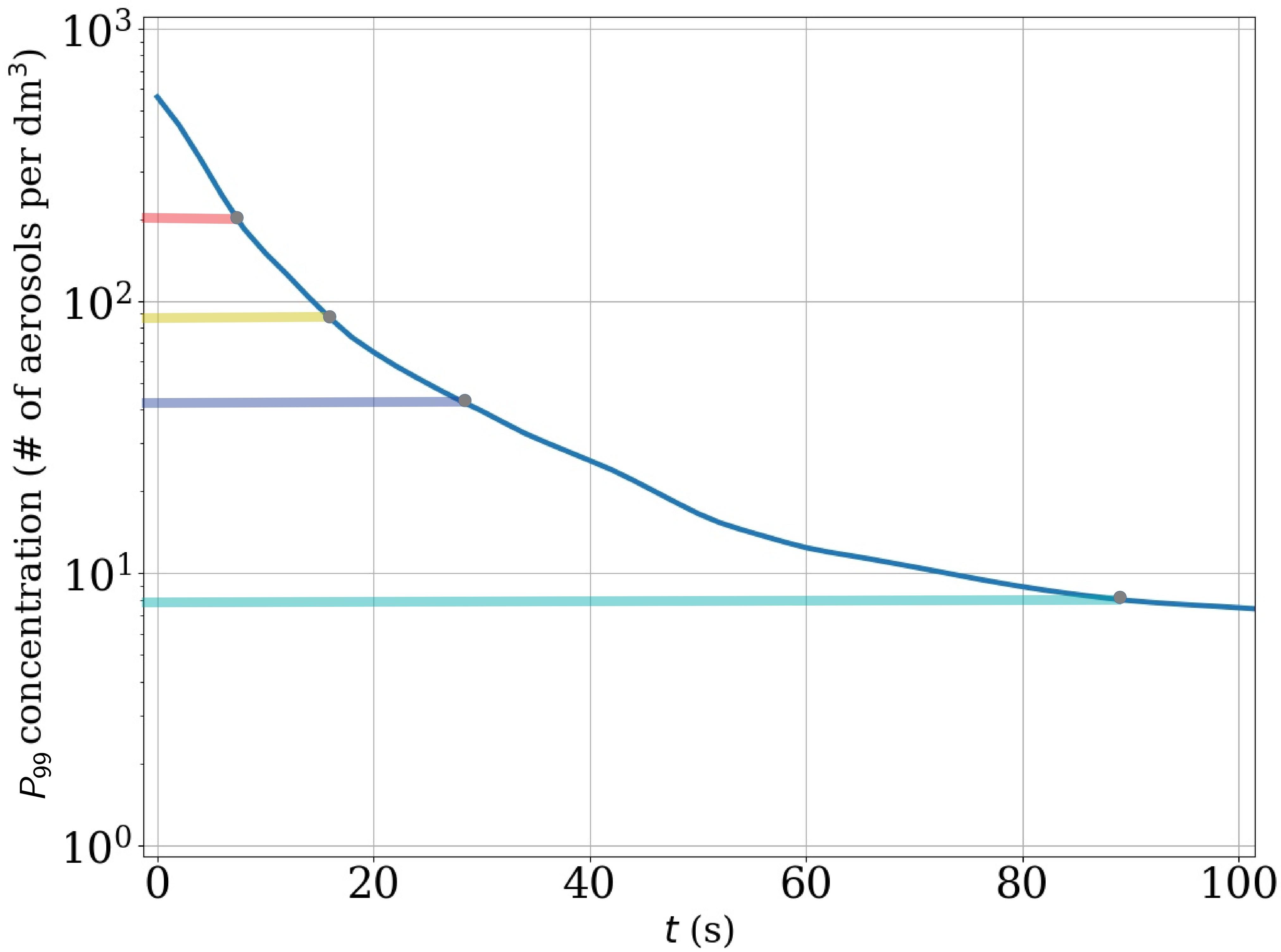}
	}
	\caption{Dilution curve of the flow realization shown in Fig.~\ref{fig:image33} including markers indicating the time instances shown in Fig.~\ref{fig:image33} a), b), c), and d). The coloured horizontal bars refer to the exposure-time bars in Fig.~\ref{fig:image33} with the same colours. For example, the figure shows that concentrations levels stay approximately $90~\unit{s}$ on a level where critical exposure could be obtained within less than $42~\unit{s}$ residence time (cyan indicator bar).}
	\label{fig:dilution_again}
\end{figure}

Although only four dilution and exposure times are shown here as an example, we note that the required dilution time is a continuous function of the selected exposure time and probability level. If 10\% probability level would be considered for the indicated exposure times, these dilution times would shorten to: $5~\unit{s}$, $11~\unit{s}$, $21~\unit{s}$ and $66~\unit{s}$, respectively. The time instances mentioned here are also shown on the dilution curve of this particular flow realization in Fig.~\ref{fig:dilution_again}. Furthermore, it is important to remember that the dimensional dilution times depend on the TKE of the background flow.  

Analyzing the time-evolving concentration histograms is informative in a probabilistic sense, but similar to the $P_{99}$ dilution curves, they do not reveal anything specific about the spatial extent of any chosen concentration level. To elaborate this, consider the volume fragments occupied by the $P_{99}$ concentration shown in a set of four 3-dimensional contour plots in Fig.~\ref{fig:image33} (center column) taken at the same time instants as the histograms. This visualization provides an instantaneous picture of the current high-exposure risk domain. But clearly, the dynamical character of the aerosol cloud complicates the determination of the spatial extent of the overall exposure risk. 

For this purpose a cumulative point of view is adopted in defining the spatial domain of elevated exposure risk. Now ,we consider the accumulated exposure of a stationary person with a breathing rate $\dot{V}_b = 0.33~\unit{dm^3\,s^{-1}}$ by evaluating a time-evolving field for the accumulated exposure $N_{\mathrm{acc}}$ from the discrete concentration dataset as follows:
\begin{equation}
    N_{\mathrm{acc}}(t_k, \mathbf{x}) 
      = \sum_i^{n(t_k)} \left< c(t_i,\mathbf{x}) \right>_{\Delta t_i} \, \dot{V}_b \; \Delta t_i 
      \label{eq:accumulation}
\end{equation}
where $n(t_k)$ is the number of time steps leading to $t_k$ and $\left< c(t_i,\mathbf{x}) \right>_{\Delta t_i}$ is the mean concentration field within time step $\Delta t_i$ expressed as number of aerosols per $\unit{dm^3}$. 
We define the domain of elevated risk (DER) as the isosurface (contour) on which $N_{\mathrm{acc}}(t,\mathbf{x}) = N_{b,\mathrm{cr}}$. This domain is visualized as a red contour in Fig.~\ref{fig:image33} (right column) for four different $t_k$ time instances.

\begin{figure}[H]
	\centering
	\makebox[0pt]{
			\includegraphics[width=\linewidth]{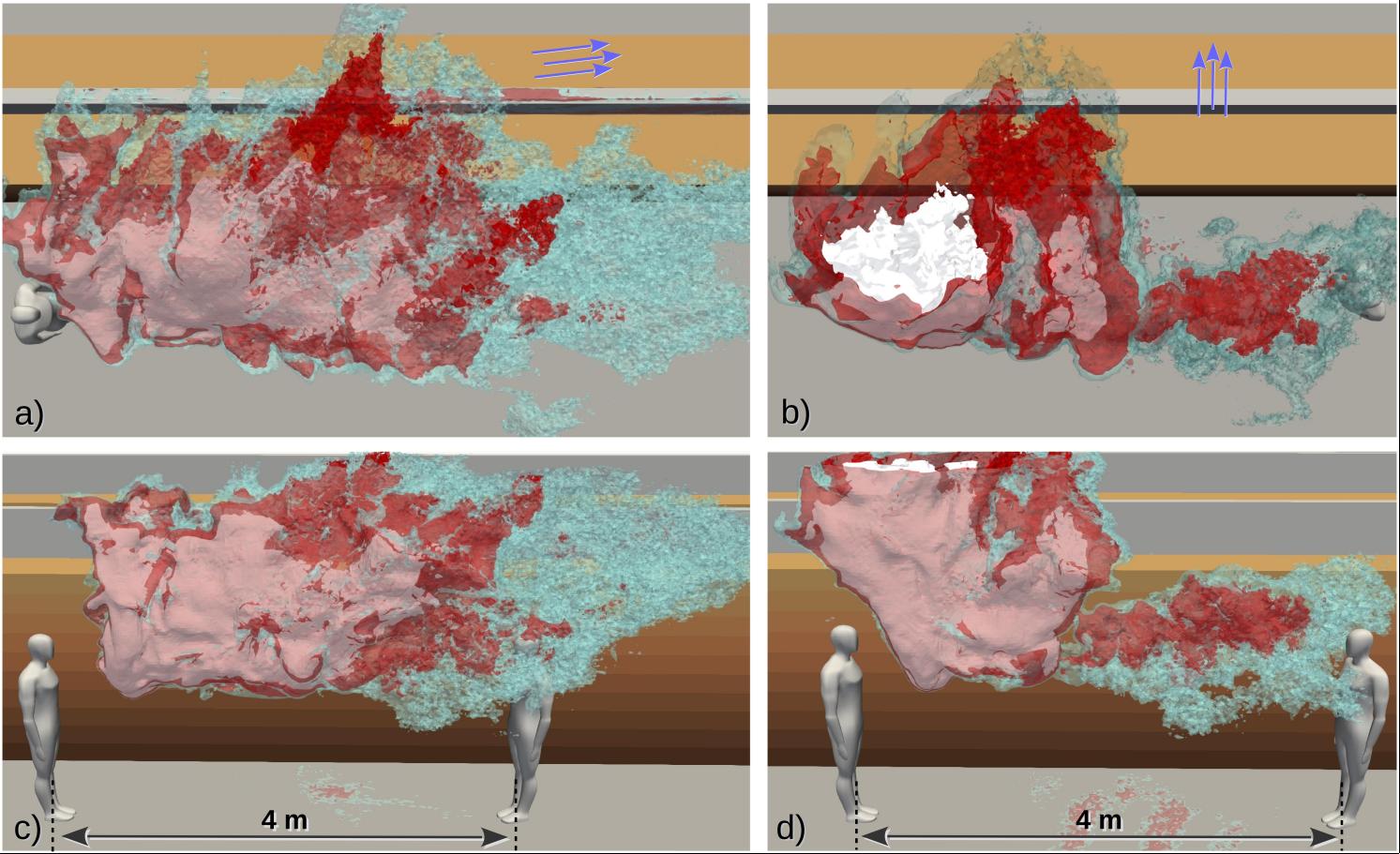}  
	}
	\caption{Visualization of the final shape and horizontal extent of the domain of elevated risk (DER) from two simulation realizations featuring a) 10~deg and b) 90~deg mean flow directions. The red contours represent the baseline DERs with $N_{b,\mathrm{cr}} = 100$ while the cyan and white contours represent the decreased $N_{b,\mathrm{cr}}=50$ and increased $N_{b,\mathrm{cr}}=150$ to study the sensitivity of the DER size to $N_{b,\mathrm{cr}}$. Note that the white contours get a rosa appearance when seen through the transparent red contours. The contours are clipped at $z=3.15~\unit{m}$.}
	\label{fig:der-extent}
\end{figure}

The images in the right column of Fig.~\ref{fig:image33} also visualize the whole cloud using three different concentration contours (blue/gray colors). In our view, the DER, being a cumulative measure, provides a better indication of the domain of risk than the instantaneous 99th or any other percentile contour. Fig.~\ref{fig:der-extent}~a) and b) illustrates the horizontal extent of two asymptotic DER realizations (red contours) where Fig.~\ref{fig:der-extent} a) depicts the final state of results shown in Fig.~\ref{fig:image33} and while Fig.~\ref{fig:der-extent}~b) is from simulation with 90~deg mean flow direction. Fig.~\ref{fig:der-extent} shows also white and cyan contours which are linked to the DER-sensitivity analysis discussed below. In both cases the DER stays in the same aisle as the coughing person and in the 90~deg case the tall shelves cause the DER to rise higher which turns out to be advantageous here. However, it should be emphasized that both cases exhibit single realizations of the turbulent flow and it is possible that in some other realization the DER might reach the neighbouring aisle.

It should be noted that a person can accumulate the critical exposure also outside the DER if he or she happens to spend sufficiently long time in locations where low concentrations of aerosols linger in the air for a long enough time. As time elapses, the concentrations decrease and the exposure times required to inhale the critical exposure at 1\% probability become larger as can be seen in Fig.~\ref{fig:image33} and in the list above. However, at the same time, volume of the DER increases and it is therefore likely that a person spends a longer time within the DER compared to earlier times with smaller DER volumes and, on the other hand, shorter required exposure times. The critical exposure can be accumulated through various paths; a person can be exposed to high concentrations for a short time, or to low concentrations for a long time, or to any combination of situations that eventually lead to the accumulation of the critical exposure. The exposure paths which individual persons present may encounter depend not only on the dispersion of the aerosol cloud but also on the behaviour of those people. The present CFD-analysis is incapable of modelling such phenomena, and therefore this problem is addressed using an event-based Monte-Carlo simulation model in Section~\ref{Sec:monte_carlo}.

Since the estimated value for the critical exposure $N_{b,\mathrm{cr}} \approx 100$ as well as the number of cough-released particles are highly uncertain, and also the breathing rates may vary significantly form person to person, it is important to study the sensitivity of the DER to these assumptions. 
Decrease of $N_{b,\mathrm{cr}}$ has the same response on the DER as the increase of number of released particles or breathing rate. 
It is sufficient to demonstrate sensitivity to $N_{b,\mathrm{cr}}$ by decreasing and increasing it by 50\% from the above used baseline value $N_{b,\mathrm{cr}} = 100$. The resulting DER contours are shown in Fig.~\ref{fig:der-extent} together with the red baseline DER-contour using cyan (decreased $N_{b,\mathrm{cr}}=50$) and white contours (increased $N_{b,\mathrm{cr}}=150$). Clearly, the variation of $N_{b,\mathrm{cr}}$ changes the spatial extent of the DER but the sensitivity is not particularly high. In our view the sensitivity is sufficiently limited to conclude that, for the studied cough phenomena, the characteristic length scale of the DER is approximately four metres.
However, if $N_{b,\mathrm{cr}}$ turns out to be very different from our estimate, then these particular DER results must be revised. This is merely a post-processing step which can be made without performing any new CFD-simulations.

\subsection{Monte-Carlo Models}
\label{Sec:monte_carlo}

We study mostly individuals walking at the regular speed of 1m/s indoors while we also discuss the low velocity limit. By model A, we assess generic public places while by model B, we discuss the particular case of a supermarket. In the end of Section~\ref{Sec:monte_carlo} we use model A to investigate indoor spaces where people are almost stagnant or move at a very low speed. We note that the numeric values of $\lambda$ and $\lambda_c=p_c N_c$ vary from person to person while being also affected by the voice loudness \citep{Asadi2019}. Our qualitative conclusions are not sensitive to these numbers. In particular for those cases with $\lambda_c=0$ the result can be directly rescaled. For $\lambda_c > 0$ a re-assessment could be easily made by the openly shared codes. And, for the theoretical exposure Eq.\ref{eq:pave}, an assessment for an average person could be made directly for any value of $\lambda$ and $\lambda_c$. 

\subsubsection{Model A: Generic Public Place - Regular Walking Speed} 
\begin{figure}[H]
\centering
  \includegraphics[width=1.0\linewidth]{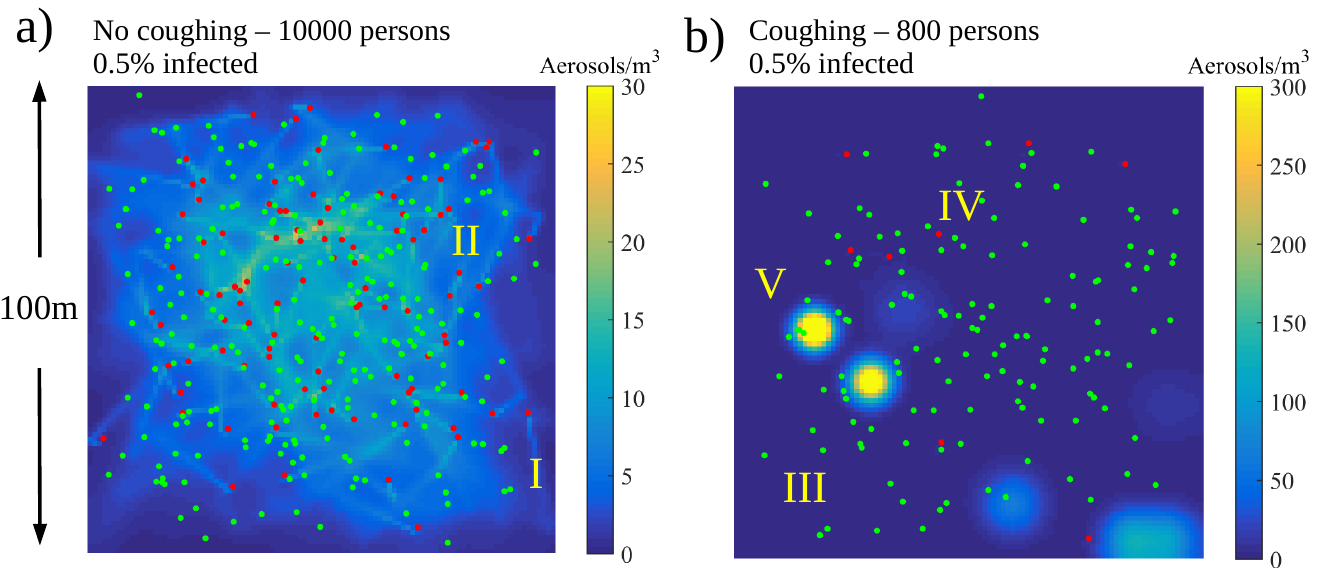}
\caption{Monte-Carlo model A based estimates on aerosol particle number density in a generic 100m by 100m area at $t=60$ minutes when 0.5\% of the population is infected. a) 10,000 persons in the area - no coughing persons. b) Only 800 persons reside in the area - coughing persons. All infected are shown (red) while only part of the susceptible persons (green) are shown for better visibility.}
\label{fig:MCdemo}
\end{figure}

Fig.~\ref{fig:MCdemo} illustrates the instantaneous aerosol particle number density ($[c]=\unit{m^{-3}}$) in the studied 100m by 100m area when 0.5\% of the persons are infected. \footnote{The numeric value would translate approximately to the number of inhaled particles during a 1h period assuming $1~\unit{m^3}$ air consumption.} Panel a) shows that in tightly spaced premises at average $N/A=1\unit{m^{-2}}$, without coughing persons, residence times of several hours can expose significant part of the individuals to $O(10)-O(50)$ aerosol particles. We note that the number would be $O(100)-O(500)$ if the value $\lambda = 50 \unit{s^{-1}}$ would have been used instead \citep{Asadi2019}. Weak aerosol wakes behind an infected individual are noted (I) along with zones of low background aerosol level (II). In fact, the motion of the persons promotes spreading of the aerosol in the system decreasing the local concentration levels. Panel b) illustrates a situation with rather sparsely populated area of $N/A=0.08\ \unit{m^{-2}}$ with coughing individuals. where the typical background concentration is respectively weaker (III). Casual encountering of the infected with the susceptible individuals remain short due to the swift motion and fast by-pass time (IV). However, when someone coughs, domains of elevated risk (DER) appear around the yellow cough plumes (V) and certain individuals are prone to radically high aerosol levels, here $600~\unit{m^{-3}}$ \footnote{Note: the scale has been truncated at 300 $\unit{m^{-3}}$ for higher contrast. Also, we do not incorporate intermittent stopping in the models which could expose certain individuals even more.}. The individuals close to the coughing persons would certainly be the most exposed ones. In Fig.~\ref{fig:MCdemo}~a) a person, who would stand still near the center of the domain, could reach the critical level $N_{b,cr}=100$ in 3-4 hours from the speech generated aerosol while the critical exposure would be reached in just 10-30 minutes for $\lambda=50s^{-1}$. Instead, in Fig.~\ref{fig:MCdemo}~b) only a few individuals would be exposed to cough plumes but the situation would be much worse if the population density would be higher or if the persons would be standing still. All in all, this example illustrates the importance of avoiding long residence times at crowded places, emphasis on efficient ventilation design to remove the aerosol as well as quarantine of the infected individuals.

\begin{figure}[H]
\centering
  \includegraphics[width=1.0\linewidth]{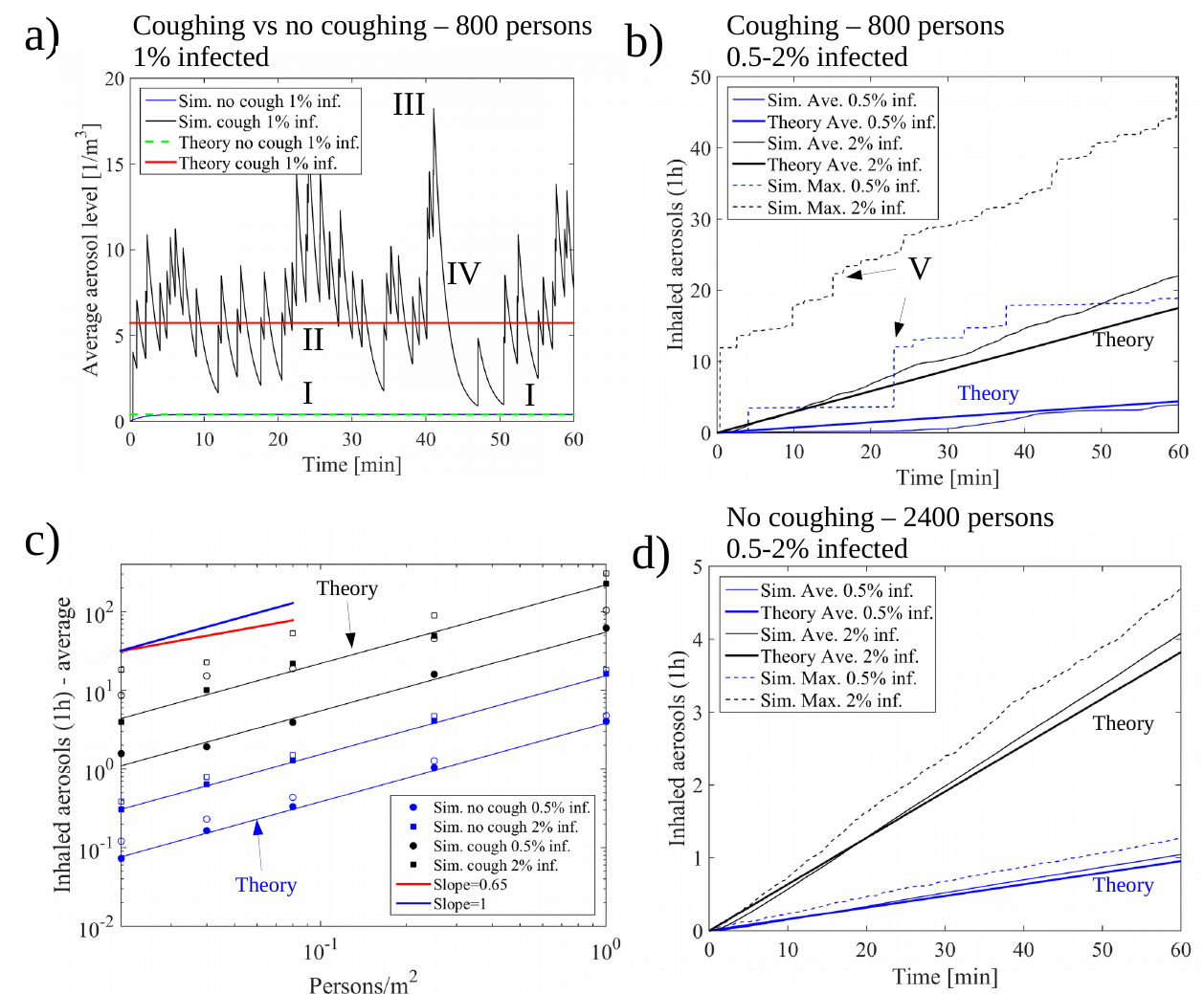}
\caption{a) Average aerosol levels in the premise follow the theoretical predictions of Eq.~\ref{eq:cave}. b) Average exposure increases linearly in time following rather well the theory. Intermittent coughs expose certain individuals. c-d) Even in asymptomatic cases, aerosol exposure of $\sim 5$ per hour is possible for large $N/A$. Under the presence of symptomatic individuals, all crowd densities pose an increased risk. }
\label{fig:MCdata}
\end{figure}

Fig.~\ref{fig:MCdata}~a) demonstrates how the average particle concentration level of the system $c_{ave}$ varies in time. Without coughing persons, the exact theoretical prediction of Eq.~\ref{eq:cave} for $p_c=0$ is trivially recovered by the simulation model (I). With coughing individuals, it is noted that $c_{ave}$ fluctuates around the theoretical line (II) predicted by Eq.~\ref{eq:cave} for $p_c=6/3600$. When someone coughs, $c_{ave}$ increases to a peak value (III) followed by an exponential decay (IV) back towards the $p_c=0$ state (I). 

Fig.~\ref{fig:MCdata}~b) demonstrates how the number of inhaled aerosols accumulates in time for an average and the most exposed individual. We note that the linear trend and the average exposure can be rather well captured by Eq.~\ref{eq:pave}. However, the average person is not representative to the most exposed individuals who encounter cough plumes time after another (V). We note that by reducing the time spent in a gathering by 50\%, an average person would be 50\% less exposed to aerosol particles by inhalation as seen also from Eq.~\ref{eq:pave}. Without coughing persons, and with even 2400 persons in the same area, the 1h exposure is on an order of magnitude lower level as revealed by Fig.~\ref{fig:MCdata} c) and d). The most exposed and the average person are within 20\% from one another while the theoretical prediction ($p_c=0$) is very close to the simulated average exposure. 

Fig.~\ref{fig:MCdata} c) illustrates that the accumulated aerosol during 1h, inhaled by the most exposed individual, scales linearly with $N/A$ i.e. number of individuals in the premises per floor area. Theoretical prediction by Eq.~\ref{eq:pave} for an average person is in very good correspondence with the simulations for $N_{b}$ as noted in Fig.~\ref{fig:MCdata}~d). All in all, these observations, supported by Eq.~\ref{eq:pave}, highlight the importance of 1) quarantine of the infected individuals, 2) social distancing and minimization of time in crowded areas with others, 3) avoiding in particular indoor places with large $N/A$ and loud speaking (high $\lambda$). The developed simple theoretical equations turn out to be predictive in terms of the average exposure of the considered population while the 2d approach offers information on the person to person exposure variation and the most extreme cases along with DER's around the cough plumes. 

\subsubsection{Model B: Supermarket - Regular Walking Speed}
\begin{figure}[H]
 \begin{center}
 \includegraphics[width=\textwidth]{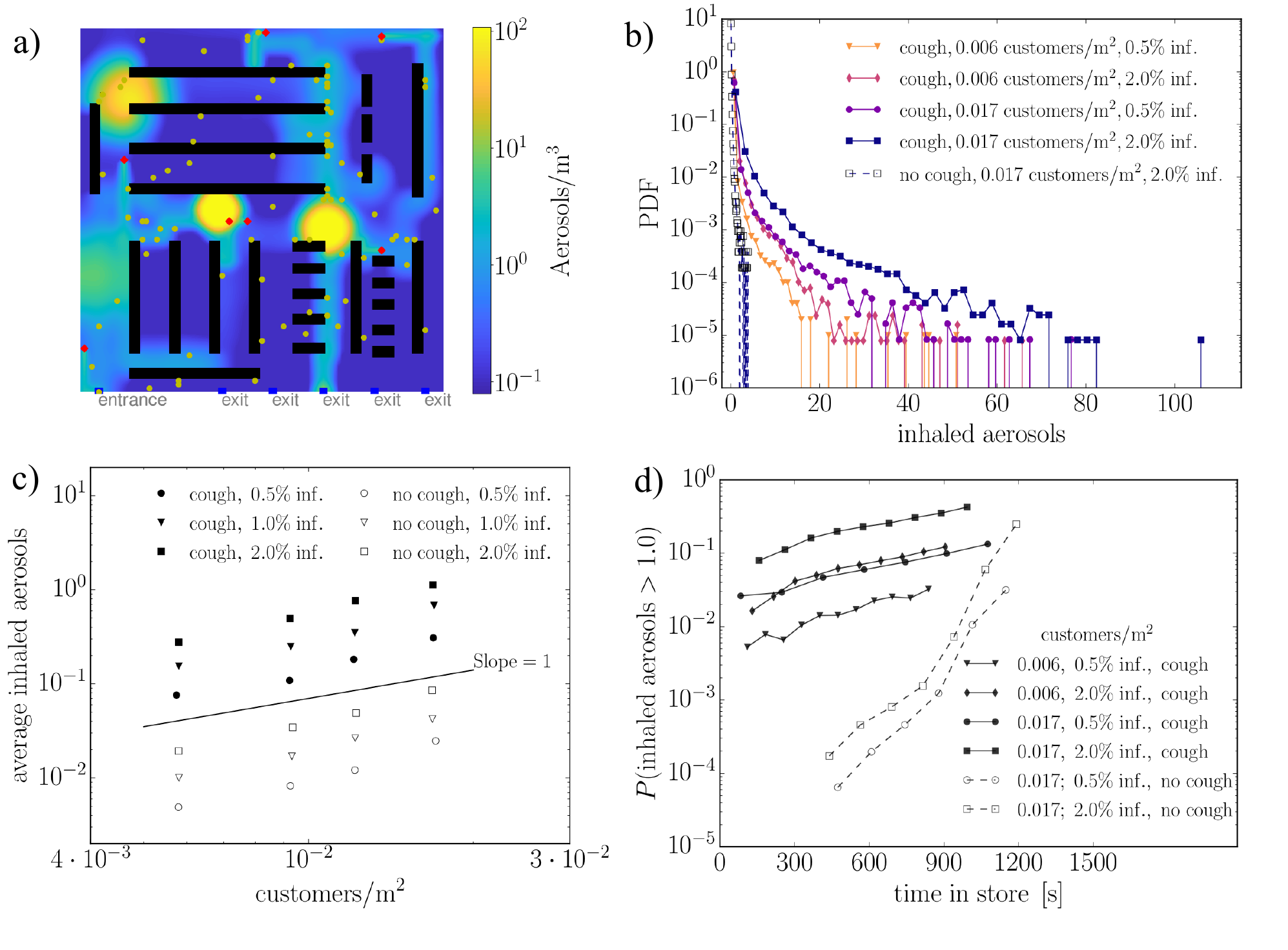}
 \caption{\label{supermarket} Monte-Carlo Model B -- supermarket case study. a) A snapshot of a typical simulation showing the healthy (yellow) and infected (red) customers walking along the aisles between the shelves. Entrance: lower left corner. Exit: lower right corner. Colorbar: aerosol concentration ($[c]=\unit{m^{-3}}$). b) The probability density function of the number of inhaled particles in coughing and non-coughing cases a at different $N/A$. Note the maximum tail probabilities in the order of $10^{-4}$ may promote significant exposure on a large population and repeating visits on a timescale of a month. c) The average inhaled aerosols as a function of the customer density. d) The probability of obtaining more than one aerosol particle as a function of time spent in the supermarket.}
 \end{center}
 \end{figure}

As another example, we study a generic supermarket. The simulation represents a floor plan of a supermarket section based on publicly available data on a typical Finnish supermarket. The system involves aisles and free spaces separated by shelves as depicted in Fig.~\ref{supermarket} a). The customers follow their personal shopping lists to gather a varying number of items \citep{Larson2005}. The customers enter the supermarket through a single entrance in the lower left corner while they exit via multiple cashiers on the lower right corner. The dynamics of movement for each customer follows a version of the shortest path algorithm ("go and get the next item"). 

In Fig.~\ref{supermarket} b) the probability density function (PDF) of inhaled aerosol particles is plotted for a statistics of 95,000 customers with varying $N/A$ and percentage of the infected customers. Additionally, a situation without coughing customers is investigated. The PDF demonstrates that the number of inhaled aerosols is very small for all the customers if coughing persons are not present. When the infected are also coughing, the probability of inhaling significant amount of aerosols is increased up to tens of particles, peaking during rush hours up to a hundred. Next, we focus on the average aerosol exposure. Fig.~\ref{supermarket} c) shows a moderately super-linear correlation between the average exposure and customer density ($N/A$), irrespective of coughing, as the supermarket throughput is limited at high $N/A$.

Finally, in Fig.~\ref{supermarket} d) we examine the exposure based on the time spent in the supermarket. The probability of customers inhaling more than 1 aerosol particles increases rapidly with the residence time (note: the y-axis is logarithmic). Surprisingly, when there are no coughing customers, the increase in the probability is more rapid, while the overall probabilities remain below the cases with coughing customers. This most likely relates to the spatial aerosol distribution. Cough plumes influence only a small portion of the customers even if the time spent in the premises increases. Instead, continuous release of aerosol becomes more homogeneously distributed involving most of the customers. While this small homogeneous background contribution is present in both cases, in the cases involving coughing it is shadowed by the large exposures of the few customers passing through the localized cough DER's.

 This study shows that the exposure during a supermarket visit even with relatively high customer densities would be quite low on a daily basis for a given individual. Coughing from active virus spreaders increases the exposure of the few unlucky individuals who happen to walk through the resulting aerosol plumes. However, despite the relatively small probability to obtain the exposure of $O(10)$ aerosol particles, considering the amount of supermarket visits daily, the number of exposed individuals can become significant in large populations. Also, we note the noted large literature uncertainty in $\lambda$ and $\lambda_c$ along with individual variations as noted by \citet{Asadi2019}. Based on this particular case study, the average probability of significant exposure during supermarket visits can be kept relatively small by 1) limiting the frequency and duration of the visits, and 2) avoiding rush hours. We also note that 3) the time spent at the cashier should be curtailed.
 
\subsubsection{Model A: High exposure risk scenarios with pre-symptomatic individuals}
During the COVID-19 pandemic, the disease has been frequently noted to transmit in somewhat longer term indoor gatherings even without symptomatic individuals. Model A can be easily utilized to assess the matter by adjusting the walking speed of the individuals to very low values $U = 0.01,\,0.1$ and $0.5~\unit{m\,s^{-1}}$, corresponding to a situation where persons stay indoors close to one another while they drift very slowly (36,360, and $1800~\unit{m}$) during the $1~\unit{h}$ time window. The lowest speed could correspond to a typical situation encountered at many work places, schools, theater, or bars. We assume that the individuals do not cough ($p_c=0$) while they produce aerosol at the same rate as previously ($\lambda = 5~\unit{s^{-1}}$). 
  
\begin{figure}[H]
 \begin{center}
 \includegraphics[width=\textwidth]{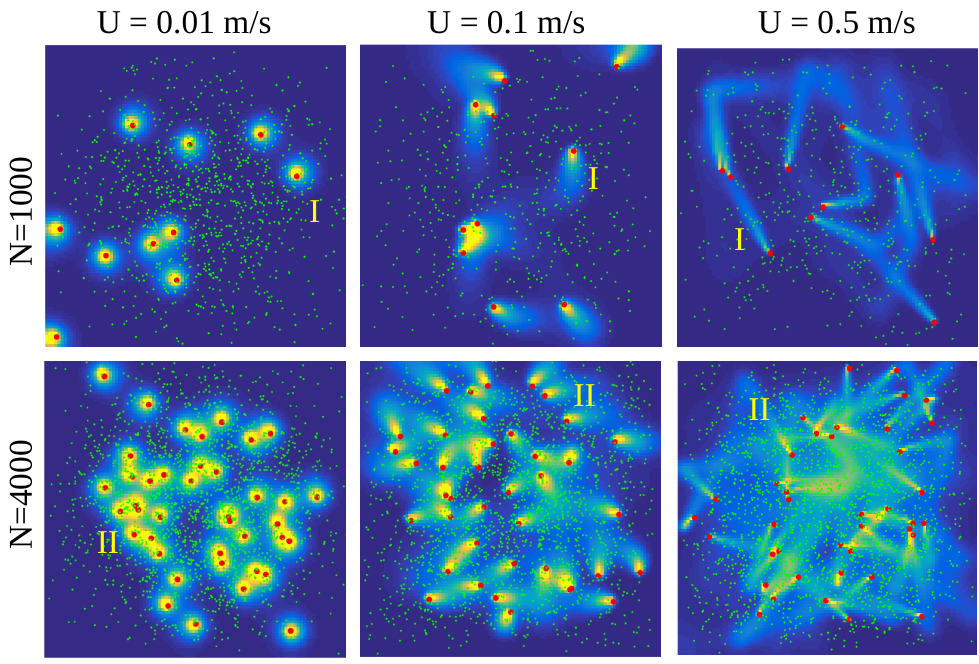}
 \caption{Monte-Carlo Model A - Comparison of very slow, slow, and intermediate drift speed without coughing individuals and $q=1\%$. While an average person is exposed almost the same for $N$ fixed, the near-stagnant motion turns out to enable very high exposure for persons near the infected. }
 \label{fig:slowmotion}
 \end{center}
 \end{figure}
 
Fig.~\ref{fig:slowmotion} shows how the aerosol cloud concentration patterns change for different configurations. The spreading patterns are qualitatively as expected since this is essentially a very well understood convection-diffusion problem with a sink and source terms in Eulerian-Lagrangian sense \footnote{One of the key non-dimensional parameters determining the spatial spreading of the concentration distribution is the Peclet number $Pe=U\delta/D$, where $\delta$ is a length scale.}. We see how the wakes of the infected persons are diffused over a broader region the faster the walking speed (I). An important aspect is  the typical by-pass time $\tau_{bp} \sim \delta/U$. Clearly, when $U$ increases, $\tau_{bp}$ decreases shortening the exposure as seen from Eq.~\ref{eq:pave}. Out of these two examples, the worst situation arises at $N=4000$ when the persons are almost stagnant (II) and close together. Then, it becomes likely that certain individuals remain a long period of time close to the infected persons. 

In fact, the probability density functions of aerosol exposure for $U =0.01~\unit{m\,s^{-1}}$ cases (not shown herein for brevity) pose flat upper tails i.e. at low velocities the most exposed individuals, close to the infected persons, are much more exposed than an average person. The higher the speed, the more fast decaying the tails become i.e. at higher velocities the most exposed individuals are not much more exposed than an average person. It is interesting that for all the speeds, for $N/A$ fixed, the average exposure in the simulations ($1~\unit{h}$) differ very little matching very well with the theoretical prediction, Eq.~\ref{eq:pave}. On its part, this example illustrates that at work places, where people stay nearly stagnant for 8-10 hours a day, the risks of getting infected may become much higher than in casual and fast by-pass encounters in public places. For example, schools, work places, or healthcare facilities could be made safer by focusing on shorter residence times, maximal inter-person distances, and remote working. As a remark, the simple theory behind Eq.~\ref{eq:pave} may serve as a tool to assess the safety of many premises, or their relative safety, quantitatively.  

\subsubsection{Exposure risks for the general population in a supermarket}
Both the models A and B underline the importance of social distance and the accumulated aerosol in time. They also stress the details of the aerosol emission mechanisms. An elevated risk follows from lengthy exposure in a confined space (indoors) with at least one infected person. With infection levels of one percent, this implies circumstances where there are in the order of 1 infected to every 100 persons present in the same premises. A way to look at the exposure probabilities is to integrate the risk over a longer period of time, for instance one month to assess the long-term implications. Assuming a population of the size of Finland (5.5~M) an order of magnitude estimate is that this will correspond to roughly $O(10^7)$ supermarket visits during such a time span. With a high exposure probability of $10^{-3}$ per visit (see again Fig.~\ref{supermarket} d)) this results in altogether $O(10^4)$ visits with high aerosol exposure. Notably, we assume that some of the individuals cough. And, aerosol exposure alone would not necessarily indicate infection. 
If there would be no coughing individuals at all, the risks of getting exposed by inhalation would be reduced. Again, this information is all contained in Eq.~\ref{eq:pave} from which a critical exposure time can be estimated for a given threshold $N_b$. A possible metric for a critical exposure time in aerosol based transmission could be: 
\begin{equation} \label{eq:tcrit}
  t_{crit} = \frac{N_{b,cr}}{c_{ave} \dot{V}_b}. 
\end{equation}  
 
\noindent A pragmatic way of looking Eq.~\ref{eq:tcrit} is via $N/A$ (see Eq.~\ref{eq:cave}): by ensuring $N/A$ is low enough and that the physical distance $\sim (N/A)^{-1/2}$ is large enough, $t_{crit}$ increases i.e. one can spend longer times in those premises. Also, Eq.~\ref{eq:tcrit} could offer a way to relate the critical exposure times of different premises, with different $N/A$, to each other 
($N_{b,cr}$ and $\dot{V}_b$ cancel out).
 
\subsection{Remarks on aerosols and ventilation}

In the PALM and OpenFOAM simulations, the ambient flow fields involved prescribed mean flow velocities over the domain of interest. To investigate the local ventilation arrangements, we used FDS to simulate aerosol transport in a domain consisting of two aisles, one below a mixing ventilation device causing a downward flow to the region of a coughing person, and another one having a very low ventilation velocity. Exhaust ventilation was then added to the ceiling, and two exhaust flow rate levels were studied: 0.18 and $1.8~\unit{m^3\,s^{-1}}$ \footnote{The smaller value corresponds to the typical air removal rate in Finnish supermarkets, $2~\unit{dm^3\,s^{-1}}$ per floor $\unit{m^2}$.}. In the beginning of the simulation, clouds of aerosol species are released in the Eulerian framework representing particles that follow the flow ($St\ll 1$).

\begin{figure}[H]
	\centering
	\includegraphics[width=\textwidth]{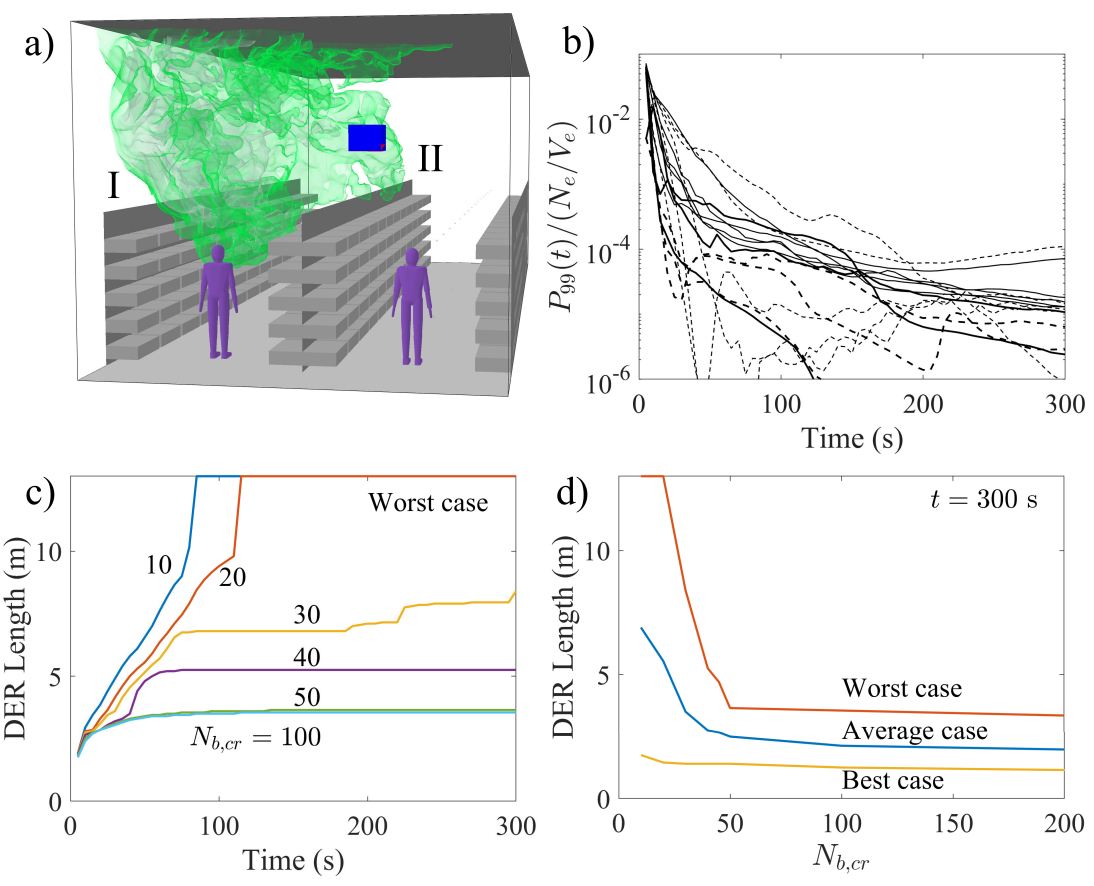}
	\caption{FDS simulation of cough clouds in a two-aisle domain. a) Cough cloud at $100~\unit{s}$ after its release on the left aisle (low mixing velocity). b) 99th percentile concentrations at height $1.6~\unit{m}$, normalized with the particle number concentration of the original cough. Dashed lines for the left aisle, solid lines for the right one. c) Maximum length of the DER among all the ventilation scenarios as a function of time for the different values of $N_{b,cr}$. d) Final ($t = 300~\unit{s}$) DER lengths for the worst, average and best ventilation scenarios. }
	\label{fig:FDS_dilution}
\end{figure}

Fig.~\ref{fig:FDS_dilution} a) illustrates the aerosol cloud $100~\unit{s}$ after a cough in the aisle with low mixing. The plume rises upward due to the prescribed exhaust rate and also buoyancy (I) along with the suction caused by the mixing ventilation device (II), which can transport part of the aerosol back to the breathing level. Fig.~\ref{fig:FDS_dilution}~b) shows the 99th percentile values of the aerosol concentrations from the different ventilation scenarios, normalized with the highest aerosol concentration $N_e/V_e$ at the origin of the cough.\footnote{Note the different normalization from the one used in Fig.~\ref{fig:image23}.} The aerosol concentrations were collected from a horizontal plane at the height $1.6~\unit{m}$.
The slowest dilution was obtained with $0.18~\unit{m^3\,s^{-1}}$ exhaust with the mixing ventilation switched off. Adding the mixing device creates a downward flow that effectively dilutes the aerosols that were released on the aisle below the device. These aerosols are transported along the aisle, decreasing the normalized concentration to the level $10^{-4}-10^{-3}$ (solid lines) further away from the cough. The dashed lines show the concentrations after a cough on the left aisle. As seen in Fig.~\ref{fig:FDS_dilution} a), the plume quickly escapes upwards, mainly due to the exhaust ventilation but also partially due to the buoyancy of the exhaled air close to the breathing zone. The aerosol particles can also return to the breathing zone after about $100~\unit{s}$ due to large scale convective air flow structures.

Overall, the most effective reduction of the concentration was obtained when the mixing ventilation was turned off, and the high exhaust was used. The reason for the good performance was the lack of horizontal mixing, allowing undisturbed buoyancy flow and effective removal of the aerosols from the breathing zone. However, the practical feasibility of such an arrangement may be questionable.

The exposure risk by inhalation is assessed by calculating the accumulation of aerosol particles in the solution points around $z = 1.6~\unit{m}$, using Eq.~\ref{eq:accumulation}. Consistent with the earlier analysis, we assumed $N_e = 40,000$ and $\dot{V}_b = 0.33~\unit{dm^3\,s^{-1}}$. The domain of elevated risk (DER) was then determined as a region where the $N_{acc}>N_{b,cr}$. As the DER was largely constrained by the aisle geometry, the physical size of the DER is here presented by its length. At each time, we can identify a 'Worst case' as a ventilation scenario having the largest DER length. Fig.~\ref{fig:FDS_dilution}~c) shows these lengths as a function of time for different $N_{b,cr}$. When $N_{b,cr}=50$ or $100$, the DER length increases from zero to its final value ($3.5~\unit{m}$) in less than $50~\unit{s}$. This result is consistent with the PALM simulations where a $4~\unit{m}$ DER was observed. Decreasing $N_{b,cr}$ to 30 gradually increases the DER length, but a drastic change is observed at $N_{b,cr}\leq20$, where the DER spans the whole aisle. Last, the final ($t=300~\unit{s}$) DER lengths of the worst, average and best ventilation scenarios at different values of $N_{b,cr}$ are shown in Fig.~\ref{fig:FDS_dilution}~d). We note that these results could be also easily re-scaled for other cough data without repeating the simulations \citep{Lindsley2012}. In summary, the provided numerical evidence indicates that significantly higher exhaust ventilation rate can efficiently promote dilution. Also, the provided example indicates how the critical exposure  $N_{b,cr}$ may affect the DER in the studied ventilation scenario. 


\section{Conclusions}
\label{sec:conclusions}

In this work we have modeled physical processes related to aerosol dispersion in air and focused on transmission by inhalation in the context of COVID-19. In large scope, such modeling efforts have relevance to the build-up of statistical metrics such as $R_0$. By May 2020, the general consensus starts to be that the SARS-CoV-2 virus does spread, not only by contact transmission, but also by inhalation of large enough numbers of airborne viral particles. In fact, the study by \citet{Doremalen2020} incidates that the SARS-CoV-2 virus may remain infectious as an aerosol for at least 3 hours. Since the concepts of droplet transmission and airborne transmission are ambiguous and person/discipline dependent, we propose that the concept 'transmission by inhalation' would better serve the purpose of communicating how infectious diseases could spread. 

Considering previous literature, we found that typical cough or speech generated droplets (e.g. $d<20~\unit{\mu m}$) are  certainly airborne from the viewpoint of physics. Even in the upper limit, an example on initially $20~\unit{\mu m}$ droplets was provided which can linger in still air for $20~\unit{min} - 1~\unit{h}$ as $\tau_{evap}\ll t_s$. We showed that droplets with diameter up to $50~\unit{\mu m} - 100~\unit{\mu m}$ could stay airborne for approximately $3~\unit{min}-20~\unit{s}$ due to the rapid drying so that they could be inhaled. In practice, air flows may carry such aerosols for much longer times but that time depends on the premise in question. The exact upper cutoff size depends on the droplet chemical composition and air RH as well. We note the large general discrepancy in the literature on droplet numbers emitted during coughing and speaking. The most important consequence of the synthesized literature and numerical observations is that, with high level of certainty, it can be concluded that a major part of droplets of respiratory origin stay airborne for long enough time for them to be inhaled. This is true even without assuming droplet drying, and particularly true when the drying is taken into account. 

Considering the inhaled aerosol amount in the analysed gatherings reported in the media, we estimated that the individuals in those occasions may have inhaled $O(100)$ aerosol particles. We emphasize that the number is an estimate and that it could vary greatly simply due to different airflow patterns in different premises or person-to-person $\lambda$ variation. Also, we acknowledge that in practice, it is possible that an individual accumulates viral particles via several transmission routes during a gathering e.g. via the toilets. Here, we have focused only on transmission by inhalation and on the physical phenomena that occur outside the human body. Considering our observation on the large variance and inconsistencies in speech and cough originated particle size and number distributions, we note that the inhaled numbers could be even an order of magnitude larger than the  estimates made herein. Also, as shown by the CFD simulations, aerosols do not necessarily spread uniformly in space due to complex character of air flows. Such non-uniform spreading would certainly affect the total exposure as well. An important outcome of the analysis is that the estimated number $O(100)$ is much larger than $O(1)$ indicating that aerosol transmission should definitely be considered as an integral entity when forming a complete picture on the epidemiology of COVID-19. 

Considering the high-resolution LES simulations on cough generated aerosol cloud evolution within a generic public environment, the following conclusions
are made. The time scale of exhaled aerosol cloud dilution is determined by the generic turbulence intensity and length scale within the indoor space, but also by the characteristic turbulent features within the considered geometry. However, there is a significant variability between individual dilution realizations due to the wide temporal range of turbulent structures naturally occurring within large ventilated indoor environments. The range of this variability has been established by a limited ensemble analysis. The risk of accumulating the critical exposure, which can be obtained via various paths, may remain significant in the order of minutes after the cough. For example, high/low exposure is accumulated in the high/low concentration zones.  

As part of our analysis, we introduced the domain of elevated risk (DER) defined as a time evolving volume within which a stationary person would already have accumulated the critical exposure. Using representative estimates for the number of exhaled aerosols and for the critical exposure to virus-laden aerosols (obtained from the order-of-magnitude analysis) and a representative value for the breathing rate, the spatial extent of DER has been demonstrated to span approximately four meters from the coughing person. The sensitivity of the DER size to $N_{b,cr}$ is analyzed by varying the estimate by $\pm 50\%$. The sensitivity of DER is confirmed to be limited enough to support the conclusion about the four meter extent of the DER for this particular setup and ambient flow. Should new information concerning the estimated parameters emerge, the demonstrated analysis can readily be performed again to conform to the latest knowledge without repeating the time-consuming CFD simulations.  

Considering the Monte-Carlo simulations and the developed exposure theory behind it, we assessed transmission by inhalation in two generic public environments. For this purpose a new aerosol transport model was developed to account for the space-time variation of the aerosol concentration. The model features an analytic solution for the average aerosol concentration level in the system along with an exposure estimate (scaling linearly with time) which both match well the simulated exposure for an average person. The simulations indicate very clearly that to minimize the risk, the first thing is to remove any symptomatic individuals from public places. Second, one should focus on the safety of areas with intermediate population densities and where people spend extended periods of time e.g. work places, or schools. Crowded public transport and bars should pay special attention on the customer number densities and ventilation aspects. We emphasize the importance of physical distance and minimal time spent in those indoor places, where high level of aerosol production is anticipated due to strong breathing or loud speaking (high $\lambda$). 

Simulations of the practical ventilation arrangements resulted in a rather similar variability in dilution rates and similar DER as what were observed in the simulations assuming a generalized turbulent background flow. A closer look at the results revealed that the nominal exhaust ventilation resulted in the slowest decay of aerosol concentration. In the region of low mixing, the buoyancy of the exhaled plume effectively removed aerosols from the breathing zone by lifting them upwards. The mixing ventilation air streams in turn diluted the concentrations in the breathing zone but were also found to capture the buoyant plumes, returning aerosols from the ceiling to the breathing zone. Considering the DER sensitivity assessment, perhaps the wisest practical conclusion is to quarantine infected persons and also stay far away from anyone coughing along with minimizing the time close to such persons. Air flows are complex and often non-intuitive. Hence, the only way to be sure that a ventilation 
works as expected is to either simulate the system by CFD or via experiments. 

Our Monte-Carlo simulation findings on the supermarket case example are noted to be consistent with the relatively low number of COVID-19 cases, the supermarket chains in Finland reported in mid May 2020 (approximately 1/1000). Based on our simulations, this could confirm that most of the infected have actually stayed at home. If the physical distance is kept long enough, the risk of getting COVID-19 even during a longer period of time, in contrast to a crowded gathering, seems low for the time being. As a major take-away, we note that Eqs.~\ref{eq:pave} and~\ref{eq:tcrit} offers a practical decision making tool to assess the "big picture" regarding the critical exposure time for the average population. According to those equations, reduction in $N/A$ by factor 1/2 may increase the critical exposure time by factor 2 respectively. Similarly, increasing air exhaust efficiency by factor 2 may allow longer residence times by factor 2. These numbers should not be taken as definitive because in practice air flows are much more complex and only a detailed assessment by CFD would give more insight to the actual times and distances. But, effectively, the proposed equations would certainly offer quantitative metrics to relate aerosol exposure in different public places to one another. In pragmatic sense, the good news behind Eqs.~\ref{eq:pave} and~\ref{eq:tcrit} is that by various common-sense actions (minimize time, maximize distance, improve ventilation,...), we could potentially reduce the exposure to aerosol with high expectations to reduce spreading of the SARS-CoV-2 virus as well. But the bad news is that it is hard to control transmission in small premises with several people. Perhaps
masks will finalize the job by lowering $\lambda$, hence lowering $c_{ave}$ and thus lowering $N_b$. 

Last, we note that the present work has various uncertainties starting from the viral content of droplets, the production rate of droplets, the chemical composition and size range of droplets in different respiratory modes, and their person-to-person variation. Also, there is a high demand for more detailed epidemiological research information on the case studies where transmission has occurred. Altogether, such information could be incorporated in Monte-Carlo and CFD models to further continue the present work.


\section*{Code availability}
The used software, PALM/FDS/OpenFOAM are all open source software. The used case setups and codes will be made openly available during June-August 2020. 

\section*{Acknowledgments}
This work was supported by the Academy of Finland grant numbers 314487 and 309570, and by the Scientific Advisory Board for Defense (MATINE) grant number VN/627/2020-PLM-9. We acknowledge CSC-IT center for science Ltd for offering the supercomputing resources for the present work. The Authors wish to thank all individuals and family members who have supported the work. In particular, Riikka Haikarainen, Kalle Kataila, Minna H\"oltt\"a, Esko Kauppinen, Olli Ranta, Otto Blomstedt, Cheng Qiang, Muhammad Saad Akram, Rahul Kallada Janardhan, Randy McDermott, and Marcos Vanella.

\bibliography{mybibfile}

\end{document}